\documentclass{article}
\usepackage{amsmath}
\usepackage{lineno}
\usepackage{graphicx}
\graphicspath{ {images/} }

\input{tcilatex}
\begin{document}

\begin{center}

\textbf{Replicator dynamics for the game theoretic selection models based on
state\bigskip }

Krzysztof Argasinski

\textit{argas1@wp.pl}\bigskip

Institute of Mathematics of Polish Academy of Sciences

ul. \'{S}niadeckich 8

00-656 Warszawa\bigskip

Ryszard Rudnicki

\textit{ryszard.rudnicki@us.edu.pl}\bigskip

Institute of Mathematics of Polish Academy of Sciences

ul. \'{S}niadeckich 8

00-656 Warszawa\bigskip

\begin{figure}[h!]
\includegraphics[width=10cm]{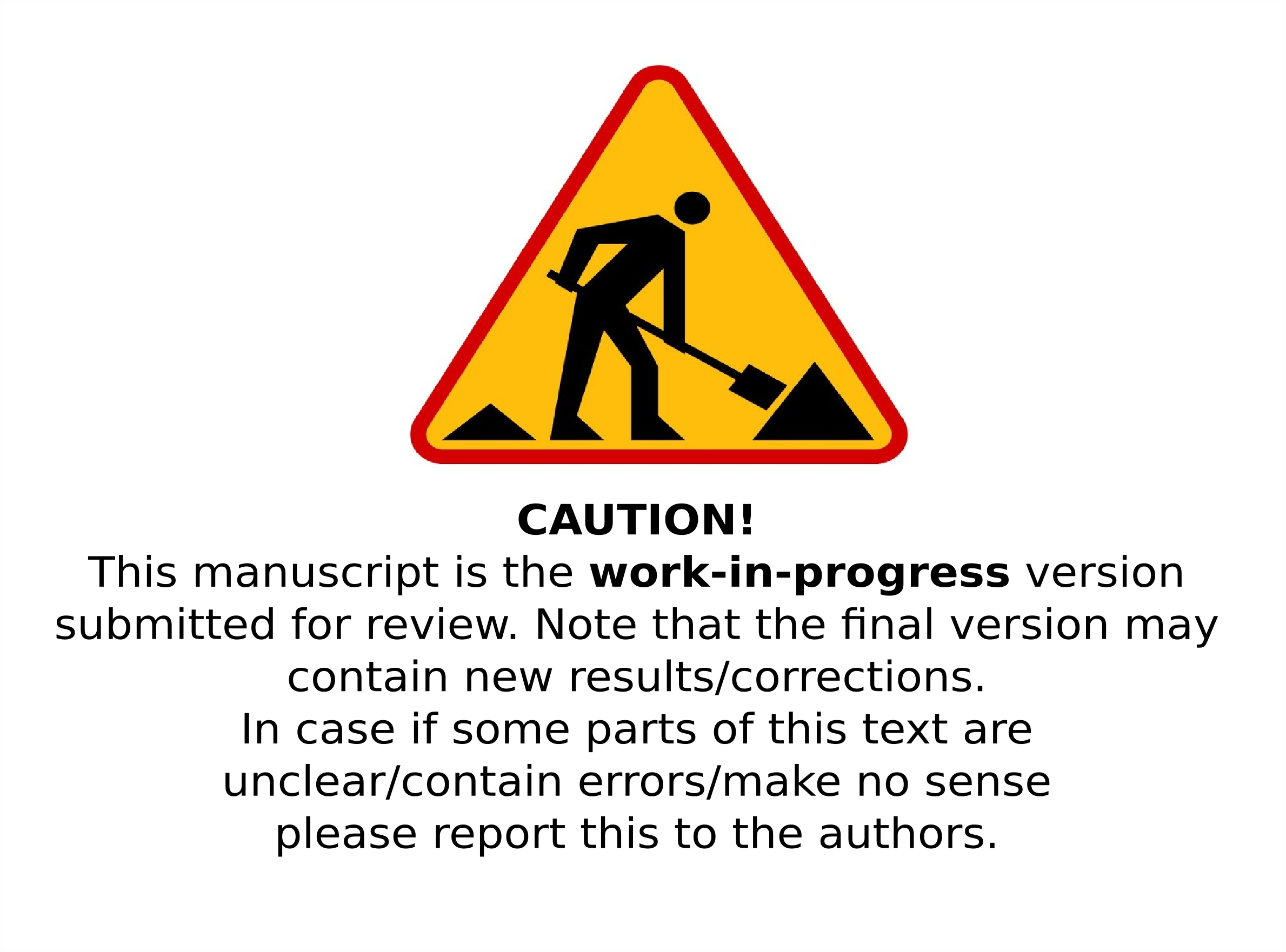}
\centering
\end{figure}

\end{center}

\textbf{keywords}: replicator dynamics, state based models, evolutionary
game, stage structured population, age structured population, Owner-Intruder
game\newpage 

\textbf{Abstract:}\newline
The paper contains the attempt to integration of the classical evolutionary
game theory based on replicator dynamics and the state based approach of
Houston and Mcnamara. In the new approach, individuals have different
heritable strategies, however the individuals carrying the same strategy can
differ on the state, role or situation in which they act. Thus, the

classical replicator dynamics is completed by the additional subsystem of
differential equations describing the dynamics of transitions between
different states. In effect the interactions described by game structure, in
addition to the demographic payoffs (constituted by births and deaths) can
lead to the change of state of the competing individuals. The special cases
of the new framework of stage structured models where the state changes
describe developmental steps or aging are derived. New approach is
illustrated by the example of Owner-Intruder game with explicit dynamics of
the role changes. New model is the generalization of the demographic version
of the Hawk-Dove game,\ the difference is that opponents in the game are drawn
from two separate subpopulations consisting of Owners and Intruders.
Intruders check random nest sites, and play the Hawk-Dove game with the
Owner if they are occupied. Interesting feedback mechanism is produced by
fluxes of individuals between subpopulations. Owners produce newborns which
become Intruders, since they should find a free nest site to reproduce.%
\newline
\newline
\newline
\newline
\newline

\section{Introduction\protect\bigskip}

The classical evolutionary game theory consists of the game structure
associated by replicator dynamics (Maynard Smith 1982; Cressman 1992;
Hofbauer and Sigmund 1988, 1998) This approach is mainly based on the simple
matrix games, where payoff matrices describe the excess from the average
growth rate in the population for the respective strategies. To add
necessary ecological details and to describe the models in measurable
parameters, the classical approach was expressed in terms of the demographic
vital rates (Argasinski and Broom 2013a, 2018a, 2018b; Zhang and Hui 2011;
Huang et al. 2015, Gokhale and Hauert 2016). In this approach instead of
single payoff function there are separate payoff functions describing the
mortality (probability of death during interaction) and fertility (offspring
number resulting from the interaction), in effect vital rates (birth and
death rate) are products of interaction rates describing the distribution of
interactions (game rounds) in time and demographic payoffs describing the
average outcomes of a single interaction. Those mortality and survival
payoff functions, describing the game interaction, may depend on each other
leading to the trade-offs, such as mortality-fertility trade-off function
describing the reproductive success of the survivors of the interaction
(Argasinski and Broom 2013a, 2017, 2018). The distinction between opposing
mortality and fertility forces was described as the cornerstone of the novel
mechanistic formulation of evolutionary theory (Doebeli et al. 2017),
however, the new approach is still based on a very strong simplifying
assumption. The individuals (and thus their payoffs) differ only by
inherited strategy and the individuals carrying the same strategy are
completely equivalent. Another thing is that many conflicts in nature have
no direct effect in the reproductive success or death, however they can
affect the level of supplies of the individual. Thus births and deaths are
not the only currency in which are paid payoffs in evolutionary games. The
alternative general approach to the game theoretic modelling, dealing with
the problem of non-heritable differences between individuals carrying the
same genes was introduced in Houston and McNamara (1999). In the state based
approach the individual differences caused by environmental conditions are
explicitly taken into consideration. Independently, basic replicator
dynamics models completed by state switching dynamics were introduced by
Brunetti et al. (2015, 2018). The goal of this paper is to integrate the
state based approach with the demographic approach to the dynamic
evolutionary games (Argasinski and Broom 2013a, 2017, 2018). We will derive
and analyze the general framework describing the dynamics underlying the
process of state changes and the interplay between this process and the
population dynamics. Some specific models dealing with the state or role
changes already exist in literature. For example, dynamics of pair formation
problem in Battle of the sexes (Myllius 1999), dynamics of role changes from
Owner to Intruder caused by population feedbacks (Kokko et al. 2006) and
formation of interacting pairs in games with time constraints (K\v{r}ivan et
al. 2018). However, the general game theoretic framework don't contain tools
allowing for modelling these problems. The last example (K\v{r}ivan et al.
2018) is described as going "beyond replicator dynamics", but in this paper
we want to show that replicator dynamics still can be useful for the
problems of this type and it is too early to send it on retirement.

\section{Methods}

We will extend the event based approach of Argasinski and Broom
(2013a,2017,2018) where individuals are involved in different activities
(described as different evolutionary games) and the growth rate of the
population is determined by the aggregated demographic outcomes of the
particular events (rounds of different games). Then the growth of the
population of the individuals with strategy $s$ can be presented as%
\begin{equation}
\dot{n}_{s}=n_{s}\sum_{j}\tau ^{j}\left( W_{s}^{j}-d_{s}^{j}\right)
\label{ab1}
\end{equation}%
where\newline
$\tau ^{j}$ is the interaction rate of $j$-th event (game type)\newline
$W_{s}^{j}$ is the fertility payoff (number of offspring)\newline
$d_{s}^{j}$ is the mortality payoff,\bigskip \newline
and products $\sum_{j}\tau ^{j}W_{s}^{j}$ and $\sum_{j}\tau ^{j}d_{s}^{j}$
constitute the demographic vital rates (Argasinski and Broom 2017). The
strategy $s$ will affect payoffs in some particular types of games or in
single type of game only, described as the \textbf{focal game} (for
simplicity we will assume this case throughout the paper) with payoff
functions $W_{s}^{F}$ and $d_{s}^{F}$. Then other games where all strategies
will obtain equal payoffs will constitute\textbf{\ background mortality} and 
\textbf{fertility rates}:%
\begin{eqnarray*}
W_{B} &=&\sum_{j}\tau _{B}^{j}W_{B}^{j}/\tau _{B}\text{ \ \ average
fertility per event (where }\tau _{B}=\sum_{i}\tau _{B}^{i}\text{)} \\
m_{B} &=&\sum_{j}\tau _{B}^{j}d_{B}^{j}/\tau _{B}\text{ \ \ \ average death
probability per event.}
\end{eqnarray*}%
Then equation (\ref{ab1}) can be presented as 
\begin{equation}
\dot{n}_{s}=n_{s}\left[ \tau _{F}\left( W_{s}^{F}-d_{s}^{F}\right) +\tau
_{B}\left( W_{B}-m_{B}\right) \right] .  \label{ab2}
\end{equation}

Then after adjustment of the timescale, the focal game occurrence rate can
be removed from the equation. In effect background birth and death rates
will be transformed into $\Phi =\frac{\tau _{B}}{\tau _{F}}W_{B}$ and $\Psi =%
\frac{\tau _{B}}{\tau _{F}}m_{B}$ leading to the simplified form of
equations (\ref{ab2}) where superscript $F$ is not necessary and the payoff
functions can depend on the composition of the population described by the
vector of strategy frequencies $q$\textbf{\ }where$\
q_{s}=n_{s}/\sum_{i}n_{i}$: 
\begin{equation}
\dot{n}_{s}=n_{s}\left[ \left( W_{s}(q)-d_{s}(q)\right) +\left( \Phi -\Psi
\right) \right] .  \label{ab3}
\end{equation}

Above system of equations can be transformed to the dynamics of the relative
frequencies of the strategies by change of coordinates $q_{s}=n_{s}/%
\sum_{i}n_{i}$ leading to the \textbf{replicator dynamics:}%
\begin{equation}
\dot{q}_{s}=q_{s}\left[ \left( W_{s}(q)-\bar{W}(q)\right) -\left( d_{s}(q)-%
\bar{d}(q)\right) \right] ,  \label{rep1}
\end{equation}

where $\bar{W}(q)=\sum_{s}q_{s}W_{s}(q)$ and $\bar{d}(q)=%
\sum_{s}q_{s}d_{s}(q)$. Note that background vital rates $\Phi $ and $\Psi $
vanishes from replicator dynamics. In more complicated cases, where
individuals differ not only on strategies but also on some another trait
such as for example sex (Argasinski 2012,2013,2017) we can use
multipopulation \ replicator dynamics (Argasinski 2006) where initial
population is divided into some subpopulations. Then the composition of each
subpopulation (indexed by superscript) described by frequencies $%
q_{s}^{j}=n_{s}^{j}/\sum_{j}n_{s}^{j}$ will be described by respective
replicator dynamics (\ref{rep1}). Those system of systems (\ref{rep1}) will
be completed by additional system describing the dynamics of relative
proportions between those subpopulations $g_{s}=n_{s}/\sum_{j}n_{s}$ (where $%
n_{s}=\sum_{j}n_{s}^{j}$ and $n=\sum_{s}n_{s}$) which will have similar form
to (\ref{rep1}) but expressed in terms of the excess of the average
subpopulation payoffs from average payoffs in the general population. The
last element that should be added is the equation on general population size
(scaling parameter). In effect we obtain%
\begin{eqnarray}
\dot{q}_{s}^{j} &=&q_{s}^{j}\left[ \left( W_{s}^{j}(g,q)-\bar{W}%
_{s}(g,q)\right) -\left( d_{s}^{j}(g,q)-\bar{d}_{s}(g,q)\right) \right] 
\label{multirep} \\
\dot{g}_{s} &=&g_{s}\left[ \left( \bar{W}_{s}(g,q)-\bar{W}(g,q)\right)
-\left( \bar{d}_{s}(g,q)-\bar{d}(g,q)\right) \right]   \notag \\
\dot{n} &=&n\left[ \bar{W}(g,q)-\bar{d}(g,q)+\left( \Phi -\Psi \right) %
\right] .  \notag
\end{eqnarray}%
Respective demographic payoffs can depend on each other leading to the
trade-off functions (Argasinski and Broom 2013a, 2017, 2018), for example
when reproduction occurs after mortality stage, instead of simple fertility
payoff $W_{s}$\ we should use mortality-fertility trade-off function where $%
s_{i}^{j}(g,q)=1-d_{i}^{j}(g,q)$ 
\begin{equation}
V_{i}^{j}(g,q)=\sum_{l}q_{l}s_{i}^{j}(e_{l})W_{i}(e_{l}),  \label{tradeoff}
\end{equation}%
where only survivors of the game round can reproduce. The above system can
be extended to explicit density dependence by some density dependent adult
mortality or incorporation of the juvenile recruitment survival to the
fertility payoffs (Argasinski and Broom 2013a, 2018a, 2018b). This can be
implemented by multiplication of the fertility rates $W_{s}^{j}(g,q)$ or $%
V_{i}^{j}(g,q)$ by suppression coefficient $(1-n/K)$. Suppression can be
interpreted as the juvenile recruitment survival. For constant mortality and
fertility rates this approach leads to the Nest Site Lottery mechanism
(Argasinski and Broom 2013b), where newborns introduced to the population
compete for the available nest sites. Under shortage of nest sites they form
the pool \ of candidates from which are drawn those who replace dead adults
in the released nest sites. It can work for other forms of suppression than
logistic growth (Rudnicki 2018). The availability of nest sites can be used
for derivation of the fully mechanistic growth model (Argasinski and
Rudnicki 2017, Argasinski and Rudnicki submitted). This is important because
changing juvenile survival alters the value of the reward in evolutionary
games (Argasinski and Broom 2013a, 2018a, 2018b) which may invert the
strategic situation leading to breakdown of the growth of some strategy
(Argasinski and Broom 2013a) or even stabilize the unstable invasion barrier
(Argasinski and Broom 2018b). The general impact of the density dependence
on selection deserves more attention (Da\'{n}ko et al 2018).

\section{Results}

\begin{tabular}{|l|}
\hline
$n_{s}^{i}$ \ \ number of individuals with strategy $s$ and in state $i$ \\ 
\hline
$q_{s}^{i}=n_{s}^{i}/\sum_{j}n_{j}^{i}$ \ \ frequency of individuals in
state $i$ among $s$-strategists \\ \hline
$g_{s}=\sum_{i}n_{s}^{i}/\sum_{z}\sum_{l}n_{z}^{l}$ \ \ frequency of $s$%
-strategists in the population \\ \hline
$\tau ^{j}$ \ \ is the interaction rate of $j$-th event (game type) \\ \hline
$W_{s}^{i}$ \ \ is the fertility payoff (number of offspring) of $s$%
-strategist in state $i$ \\ \hline
$d_{s}^{i}$ \ \ \ is the mortality payoff of $s$-strategist in state $i$ \\ 
\hline
$\Phi ^{i}$ \ \ background fertility rate in state $i$ \\ \hline
$\Psi ^{i}$ \ \ background mortality rate in state $i$ \\ \hline
$c_{s}^{i,z}(g,q)$ \ \ \ switching payoff (probability of transition from
state $i$ to $z$ ) \\ \hline
$c_{s}^{i}=\sum_{k\neq i}c_{s}^{i,k}=1-c_{s}^{i,i}$ \ \ \ probability of
leaving the state $i$ \\ \hline
$\Lambda ^{z,i}$ \ \ \ background switching rate \\ \hline
$\Lambda ^{i}$ \ \ \ \ \ background leaving rate \\ \hline
$V_{s}^{i}(q,g)$ \ \ \ survival-fertility trade-off function \\ \hline
$Z_{s}^{i,j}(q,g)$ \ \ survival-switching trade-off function \\ \hline
$X_{s}^{i}(q,g)$ \ \ \ switching-survival trade-off function \\ \hline
$Y_{s}^{i}(q,g)$ \ \ \ switching-mortality trade-off function \\ \hline
$\bar{W}_{s}(g,q)$, $\bar{W}(g,q)$ \ \ average fertility payoff of $s$%
-strategists/whole population \\ \hline
$\bar{s}_{s}(g,q)$, $\bar{s}(g,q)$ \ \ \ \ average survival payoff of $s$%
-strategists/whole population \\ \hline
$\bar{c}_{s}(g,q)$, $\bar{c}(g,q)$ \ \ \ \ average switching payoff of $s$%
-strategists/whole population \\ \hline
$\bar{\Phi}_{s}(q_{s})$, $\bar{\Phi}(g,q)$ \ \ \ \ average background
fertility rate of $s$-strategists/whole population \\ \hline
$\bar{\Psi}_{s}(q_{s})$, $\bar{\Psi}(g,q)$ \ \ average background mortality
rate of $s$-strategists/whole population \\ \hline
$\bar{\Lambda}^{i}$, $\bar{\Lambda}$ \ \ \ average background leaving rate
of state $i$/whole population \\ \hline
$H^{O}$, $H^{I}$ \ \ \ proportion of Hawks among Owners/Intruders \\ \hline
$\Phi _{s}^{I}=\frac{q_{s}^{O}}{1-q_{s}^{O}}\Phi $ Intruders per capita
increase rate caused by Owners background\ fertility \\ \hline
\end{tabular}

\textbf{Table 1 List of important symbols}\bigskip \bigskip

\textbf{Plan of the paper:}\newline
-completion of the framework presented in introduction by switching payoffs
describing state changes resulting from the game (sections 3.1-3.2).\newline
-introduction of the trade-off functions analogous to the
mortality-fertility trade-off function (\ref{tradeoff}) describing the
trade-offs between demographic payoffs and switching payoffs (section 3.3).%
\newline
-introduction of ratio dependent nonuniform interaction rates, when
interactions occur between individuals in different states (Males and
Females or Owners with Intruders), section 3.4.\newline
-derivation of the generalized replicator dynamics, where selection
equations are completed by equations describing the dynamics of fluxes
between different states among carriers of different strategies (sections
3.5-3.8), including the limit cases of separation of timescales between
switching and demographic dynamics (section 3.9).\newline
-special cases of the obtained framework describing the stage and age
structured populations (section 4).\newline
-example of Owner-Intruder game describing the competition for nest sites,
with explicit dynamics of fluxes between roles of settled Owner and homeless
Intruder.

\subsection{Population growth with switching payoffs}

Here we will extend the framework from the previous section to the state
dependent case. Table 1 contains list of important symbols. Therefore,
assume that individuals in the population differ in states and similarly to
the previous section are involved in the different types of events,
described by demographic outcomes (mortality and fertility). Then each type
of interaction event can be characterized by following parameters:\bigskip 
\newline
$\tau ^{j}$ is the interaction rate of $j$-th event, since it is not
affected by state.\newline
$d_{s,i}^{j}$ is the mortality payoff (probability of death during
interaction event)\newline
$W_{s,i}^{j}$ is the fertility payoff (number of offspring resulting from
the interaction)\newline
of $s$-strategist in state $i$ obtained in $j$th type event.\bigskip \newline
However, state is not heritable, offspring may have different state than
parental individual. We can assume that the state of the offspring is drawn
with the probability $\zeta _{s,i}(k)$ for $k$-th state. Then individual $%
s,i $ will produce $O_{s,i}^{j}\zeta _{s,i}(k)$ individuals in state $k$
(where $O_{s,i}^{j}$ is the mating payoff which describes the number of
offspring obtained by $s,i$ in the $j$-th type event). Then $%
\sum_{z}n_{s,z}O_{s,z}^{j}\zeta _{s,z}(k)$ individuals in state $k$ will be
produced leading to the following average aggregated per capita fertility
rate of individual $s,i$ caused by the $j$-th type of event:%
\begin{equation}
W_{s,i}^{j}=\dfrac{\sum_{z}n_{s,z}O_{s,z}^{j}\zeta _{s,z}(i)}{n_{s,i}}
\label{fertility}
\end{equation}%
\bigskip \newline
and $O_{s,i}^{j}$ is the mating payoff\ (successful mating attempts). The
general growth equation equivalent to (\ref{ab1}) of the subpopulation of
individuals in state $i$ and with strategy $s$ (described by the subscript
vector $s,i$ while superscripts will describe event type will be: \bigskip

\begin{equation}
\dot{n}_{s,i}=n_{s,i}\sum_{j}\tau ^{j}\left( W_{s,i}^{j}-d_{s,i}^{j}\right)
\label{basic}
\end{equation}%
We should add the third type of the event outcome, the probability of switch
in the state from $i$ to $k$ (for $i\neq k$)\ described by $c_{s,i,k}^{j}$
(assume that $c_{s,i}^{j}=\sum_{k\neq i}c_{s,i,k}^{j}=1-c_{s,i,i}^{j}$
describes the per capita probability of leaving the state $i$). This idea is
similar to the state switching dynamics introduced by Brunetti et al. (2015,
2018). Then the population should be divided into different state classes.
Then we should update our equation to the following form where term $%
\sum_{j}n_{i}\tau ^{j}c_{s,i}^{j}$ describes the \textbf{leaving rate}, i.e.
changes of a state of the individuals in the $i$-th state, while term $%
\sum_{j}\sum_{z}n_{z}\tau ^{j}c_{s,z,i}^{j}$ describes the \textbf{incoming
rate}, per capita increase of individuals in state $i$ caused by switches of
individuals at the other states.

\begin{eqnarray}
\dot{n}_{s,i} &=&\sum_{j}n_{s,i}\tau ^{j}\left(
W_{s,i}^{j}-d_{s,i}^{j}\right) -n_{s,i}\sum_{j}\tau
^{j}c_{s,i}^{j}+\sum_{j}\sum_{z\neq i}n_{s,z}\tau ^{j}c_{s,z,i}^{j} \\
&=&n_{s,i}\left[ \sum_{j}\tau ^{j}\left( W_{s,i}^{j}-d_{s,i}^{j}\right)
+\sum_{j}\tau ^{j}\left( \sum_{z\neq i}\dfrac{n_{s,z}}{n_{s,i}}%
c_{s,z,i}^{j}-\sum_{j}c_{s,i}^{j}\right) \right]  \notag
\end{eqnarray}

\subsection{Extraction of the focal type of interaction\protect\bigskip}

Now we can extract focal interactions affected by analyzed trait from the
general dynamics and assume that they are functions of the population
composition i.e. game payoffs. The population state is described by vector $%
(q_{s,i},g_{s})$:\bigskip \newline
$q_{s,i}=n_{s,i}/\sum_{l}n_{s,l}$ state distributions among carriers of the
particular strategies\bigskip \newline
$g_{s}=\sum_{i}n_{s,i}/\sum_{z}\sum_{l}n_{z,l}$ \ \ strategy
frequencies\bigskip \newline
See fig.1 for the schematic presentation of the phase space. Individuals
enter an focal game (with payoffs $W_{s,i}^{F}(g,q)$ and $d_{s,i}^{F}(g,q)$
where auxiliary index $F$ means "focal event") at rate $\tau _{F}$ as in
equation (\ref{basic}), and engage in other activities at rates described by 
$\tau _{B}^{i}$; we can consider a single class of all such activities, as
we show below. Assume that background events do not depend on the strategies
but are affected by state, since for example energy level will have impact
on the overall performance of the organism. Each of the background events
can be characterized by outcomes which include a fertility $W_{B,i}^{j}$ and
mortality $d_{B,i}^{j}$ (lower index $B$ means "background event").\ We
should also consider the background switching dynamics where%
\begin{equation*}
c_{B,z,i}=\sum_{j}\tau _{B}^{j}c_{B,z,i}^{j}/\tau _{B}\text{ \ \ \ and \ \ \
\ \ }c_{B,i}=\sum_{j}\tau _{B}^{j}c_{B,i}^{j}\text{,}
\end{equation*}%
because it will be affected by the actual state distribution determined by
impact of the focal game. We can calculate the outcomes of the average
background event for the individual in state $i$:%
\begin{eqnarray*}
W_{B,i} &=&\sum_{j}\tau _{B}^{j}W_{B,i}^{j}/\tau _{B}\text{ \ \ average
fertility per event (where }\tau _{B}=\sum_{i}\tau _{B}^{i}\text{)} \\
m_{B,i} &=&\sum_{j}\tau _{B}^{j}d_{B,i}^{j}/\tau _{B}\text{ \ \ \ average
death probability per event.}
\end{eqnarray*}%
In effect \textquotedblleft background events\textquotedblright\ occur at
intensity $\tau _{B}$ and individuals involved in those events obtain
fertility $W_{B,i}$ on average and die with probability $m_{B,i}$. Then
equation (\ref{basic}) can be presented similarly to (\ref{ab2}) in the
form: 
\begin{eqnarray}
\dot{n}_{s,i} &=&n_{s,i}\tau _{F}\left[ \left(
W_{s,i}^{F}(g,q)-d_{s,i}^{F}(g,q)\right) +\left( \sum_{z\neq i}\dfrac{n_{s,z}%
}{n_{s,i}}c_{s,z,i}^{F}(g,q)-c_{s,i}^{F}(g,q)\right) \right]  \notag \\
&&+n_{s,i}\tau _{B}\left[ \left( W_{B,i}-m_{B,i}\right) +\left( \sum_{z\neq
i}\dfrac{n_{s,z}}{n_{s,i}}c_{B,z,i}-c_{B,i}\right) \right] .
\label{basicmalth}
\end{eqnarray}%
We can adjust the timescale to make the focal game's vital rates equal to
their demographic payoffs. This will keep the mechanistic interpretation as
the number of offspring and the death probability during the interaction
event. It is clear that only the ratio of our two interaction rates is
important for the evolution of the population. Similarly to (\ref{ab3}),
after a change of timescale $\tilde{t}=t\tau _{F}$, $\tau _{F}$ vanishes and 
$\tau _{B}$ transforms into $\theta =\dfrac{\tau _{B}}{\tau _{F}}$. Since
demographic parameters $W_{B,i}$ and $m_{B,i}$ never occur without the ratio
between intensities $\theta $, we can simplify this by substitutions $\Phi
^{i}=\theta W_{B,i}$ and $\ \Psi ^{i}=\theta m_{B,i}$, constituting the
background vital rates. In similar way we can derive the background
switching intensities $\Lambda ^{z,i}=\theta c_{B,z,i}$ and respectively $%
\Lambda ^{i}=\theta c_{B,i}$ This leads to: 
\begin{eqnarray}
\dot{n}_{s,i} &=&n_{s,i}\left[ \left(
W_{s,i}^{F}(g,q)-d_{s,i}^{F}(g,q)\right) +\left( \sum_{z\neq i}\dfrac{n_{s,z}%
}{n_{s,i}}c_{s,z,i}^{F}(g,q)-c_{s,i}^{F}(g,q)\right) \right]  \notag \\
&&\text{ \ }+n_{s,i}\left[ \left( \Phi ^{i}-\Psi ^{i}\right) +\left(
\sum_{z\neq i}\dfrac{n_{s,z}}{n_{s,i}}\Lambda ^{z,i}-\Lambda ^{i}\right) %
\right] .  \label{gr}
\end{eqnarray}%
Since we extracted the focal interaction and averaged the background
interactions, we can simplify the notation by removing the superscript $F$,
since it is not necessary now. We can assume that from now subscript
describes the strategy while superscript describes the state. This leads to
the following general growth equation:\bigskip

\begin{eqnarray}
\dot{n}_{s}^{i} &=&{}n_{s}^{i}\left[ \left(
W_{s}^{i}(g,q)-d_{s}^{i}(g,q)\right) +\left( \sum_{z\neq i}\dfrac{n_{s}^{z}}{%
n_{s}^{i}}c_{s}^{z,i}(g,q)-c_{s}^{i}(g,q)\right) \right]  \notag \\
&&+n_{s}^{i}\left[ \left( \Phi ^{i}-\Psi ^{i}\right) +\left( \sum_{z\neq i}%
\dfrac{n_{s}^{z}}{n_{s}^{i}}\Lambda ^{z,i}-\Lambda ^{i}\right) \right] .
\label{malth}
\end{eqnarray}

\subsection{Causal structure of the focal interaction\protect\bigskip}

Following Argasinski and Broom (2012) we can describe the order of the
different outcomes of the focal interaction. For example only survivors of
the interaction can reproduce. Then we have survival payoff function $%
s_{s}^{i}(q,g)=1-d_{s}^{i}(q,g)$. To derive the mortality-fertility
trade-off function where reproductive success of survivors is described by
product of survival and fertility $s_{s}^{i}W_{s}^{i}$ (Argasinski and
Broom, 2012), outcomes of interactions with all strategies present in the
population should be averaged. Recall that the population state is described
by set of vectors $\left[ g,q\right] $ where $q$ consists of vectors of
state distributions $q_{s}$ for all strategies. Monomorphic uniform
population consisting of $s$ strategists in state $i$ , where $%
q_{s}^{i}=g_{s}=1$ and all other entries are zeros, can be described by
state vector $u_{s}^{i}=[e_{s},e_{s}^{i}]$ where $e_{s}$ is the unit vector
for argument $g_{s}$ and $e_{s}^{i}$ is the unit vector with $1$ on $i$th
place for argument $q_{s}^{i}$, state distribution vectors $q_{j\neq s}^{i}$
for other strategies are zeros. It can be used as an argument of the payoff
functions to simplify the notation. Then for example $s_{s}^{i}(u_{k}^{l})$
will describe survival of $s$ strategists in state $i$ playing with $k$th
strategy carrier in state $l$. Then the \textbf{mortality-fertility trade-off%
} equivalent to (\ref{tradeoff}) function will be:%
\begin{equation*}
V_{s}^{i}(q,g)=\sum_{k}g_{k}%
\sum_{l}q_{k}^{l}s_{s}^{i}(u_{k}^{l})W_{s}^{i}(u_{k}^{l})
\end{equation*}%
similarly when only survivors of the interaction can switch to another state
we can introduce the \textbf{survival-switching trade-off function}%
\begin{equation}
Z_{s}^{i,j}(q,g)=\sum_{k}g_{k}%
\sum_{l}q_{k}^{l}s_{s}^{i}(u_{k}^{l})c_{s}^{i,j}(u_{k}^{l})  \label{m-s}
\end{equation}%
that will replace functions $c_{s}^{i,j}$ in the equations from previous
sections. Then the function $c_{s}^{i}$ describing the probability of
leaving the state $i$ should be replaced by%
\begin{eqnarray*}
Z_{s}^{i}(q,g) &=&\sum_{j\neq
i}Z_{s}^{i,j}(q,g)=\sum_{k}g_{k}\sum_{l}q_{k}^{l}s_{s}^{i}(u_{k}^{l})\sum_{j%
\neq i}c_{s}^{i,j}(u_{k}^{l}) \\
&=&\sum_{k}g_{k}\sum_{l}q_{k}^{l}s_{s}^{i}(u_{k}^{l})c_{s}^{i}(u_{k}^{l}).
\end{eqnarray*}%
We can also imagine the situation that mortality acts after the state
switching (thus only those individuals will die which remained in the focal
state) leading to\textbf{\ switching-survival\ trade-off function} 
\begin{equation}
X_{s}^{i}(q,g)=\sum_{k}g_{k}\sum_{l}q_{k}^{l}\left[ 1-c_{s}^{i}(u_{k}^{l})%
\right] s_{s}^{i}(u_{k}^{l})=1-Z_{s}^{i}(q,g)  \label{s-s}
\end{equation}%
that will replace functions $s_{s}^{i}$ and then the\textbf{\
switching-mortality trade-off function}, equivalent to $%
d_{s}^{i}=(1-s_{s}^{i})$ will be 
\begin{eqnarray*}
Y_{s}^{i}(q,g) &=&\sum_{k}g_{k}\sum_{l}q_{k}^{l}\left[ 1-c_{s}^{j}(u_{k}^{l})%
\right] d_{s}^{i}(u_{k}^{l}) \\
&=&\sum_{k}g_{k}\sum_{l}q_{k}^{l}\left[ 1-c_{s}^{i}(u_{k}^{l})\right]
-X_{s}^{i}(q,g) \\
&=&Z_{s}^{i}(q,g)-\sum_{k}g_{k}\sum_{l}q_{k}^{l}c_{s}^{i}(u_{k}^{l}).
\end{eqnarray*}%
Note that since $d_{i}^{j}=(1-s_{i}^{j})$ we have 
\begin{equation*}
X_{s}^{i}+Y_{s}^{i}=\sum_{k}g_{k}\sum_{l}q_{k}^{l}\left[
1-c_{s}^{i}(u_{k}^{l})\right] ,
\end{equation*}%
which describes the average probability of remaining in the same state.

\subsection{Frequency dependence and the interaction rates}

Note that the respective payoff functions presented above describe outcomes
of the average focal interaction. The argument of those functions should be
vector of the population state describing distribution of states for all
strategies $q_{j}^{i}$ and the strategy frequencies $g_{i}$. This will be
enough for description of the frequency dependent selection among randomly
paired individuals with different strategies and states. However, this is
not the only case. We can imagine situation, when interactions occur only\
between individuals in different particular states (such as when owners of
the habitat interact only with the intruders). In the simplest "gas" model
of panmictic population of randomly meeting individuals this can be realized
by assumption that only interactions between those particular types have
nonzero payoffs. However, we can imagine situations when pairing is not
completely random,\ such as mating when males search for females or
owner-intruder problem when homeless individuals investigate the nest-sites.
Then random pairing is limited to drawing opponents from opposite subgroups.
Then pure frequency dependence should be completed by the impact of the
proportion between interacting subpopulations. This is important, because if
the ratio between interacting subgroups is not 1, than individuals from
different subgroups will have different chances of interactions determined
by availability of potential opponents/partners. This will result in
different average numbers of interactions for different types (this will be
shown later by Owner-Intruder example). Therefore, we cannot limit ourselves
to description of the average outcomes of interaction only. This may lead to
badly defined models, similarly to the case of bimatrix games, which are
independent of the proportion between playing subpopulations (Argasinski
2006). Different numbers of interactions for different types/states can be
described by resulting different interaction rates (Argasinski and Broom
2017). For example if we have asymmetric pairwise interactions between
individuals acting in two opposite roles (such as Owner-Intruder conflict)
and the distribution of states is described by $q_{i}^{j}$ for $it$h
strategy in the $j$th role. Then, on average, ratio of individuals in role 1
to individuals in role 2 is $\sum_{s}g_{s}q_{s}^{1}/%
\sum_{s}g_{s}(1-q_{s}^{1})$ and it is proportional to the ratio of per
capita interaction rates for both roles, since individuals in minority will
always find the opponent, while those in majority not, due to shortage of
individuals of the opposite type. Thus, for example, if for minority type we
have $\tau _{F}^{1}=1$ then for majority type we have $\tau
_{F}^{2}=\sum_{i}g_{i}q_{i}^{1}/\sum_{i}g_{i}q_{i}^{2}$. Then, for example,
the respective death rates in (\ref{malth}), and in the resulting replicator
equations, will be%
\begin{equation}
D_{i}^{1}(g,q)=d_{i}^{1}(g,q)=1-s_{i}^{1}(g,q)\text{ \ \ for \ \ role \ \ 1 }
\label{deathrate1}
\end{equation}%
\begin{equation}
D_{i}^{2}(g,q)=d_{i}^{2}(g,q)\dfrac{\sum_{i}g_{i}q_{i}^{1}}{%
\sum_{i}g_{i}(1-q_{i}^{1})}=\left( 1-s_{i}^{2}(g,q)\right) \dfrac{%
\sum_{i}g_{i}q_{i}^{1}}{\sum_{i}g_{i}(1-q_{i}^{1})}\text{ \ \ for role 2}
\label{deathrate2}
\end{equation}%
If the interaction rate decreases due to the shortage of the opponents of
the opposite type we have following interaction rates%
\begin{eqnarray}
\tau _{F}^{1} &=&\left\{ 
\begin{array}{c}
1\text{ \ \ \ \ \ \ \ \ \ \ \ \ \ \ \ \ \ \ \ for\ \ \ \ \ \ \ \ }\dfrac{%
\sum_{i}g_{i}q_{i}^{1}}{\sum_{i}g_{i}(1-q_{i}^{1})}\leq 1 \\ 
\dfrac{\sum_{i}g_{i}(1-q_{i}^{1})}{\sum_{i}g_{i}q_{i}^{1}}\text{ \ \ \ for \
\ \ \ \ }\dfrac{\sum_{i}g_{i}q_{i}^{1}}{\sum_{i}g_{i}(1-q_{i}^{1})}>1%
\end{array}%
\right.  \label{tau1F} \\
\tau _{F}^{2} &=&\left\{ 
\begin{array}{c}
\dfrac{\sum_{i}g_{i}q_{i}^{1}}{\sum_{i}g_{i}(1-q_{i}^{1})}\text{ \ \ \ for \
\ \ \ \ }\dfrac{\sum_{i}g_{i}q_{i}^{1}}{\sum_{i}g_{i}(1-q_{i}^{1})}\leq 1 \\ 
1\text{ \ \ \ \ \ \ \ \ \ \ \ \ \ \ \ \ \ for\ \ \ \ \ \ \ \ \ \ \ }\dfrac{%
\sum_{i}g_{i}q_{i}^{1}}{\sum_{i}g_{i}(1-q_{i}^{1})}>1%
\end{array}%
\right. ,  \label{tau2F}
\end{eqnarray}%
and due to previous derivation based on the change of the timescale, it will
act as the multiplicative factor on the r.h.s. of (\ref{malth}). Similarly,
difference in the interaction rates for different states/types should be
incorporated into switching payoffs constituting switching rates and the
trade-off functions constituting\ conditional fertility and death rates and
conditional switching. In the next sections we will limit ourselves to the
basic simple demographic and switching rates and focus on the dynamics. When
it is necessary, simple payoff functions will be replaced by more detailed
tradeoff functions and nontrivial interaction rates described in this and in
the previous section.\bigskip

\subsection{Derivation of the replicator dynamics}

\begin{figure}[h!]
\centering
\includegraphics[width=12cm]{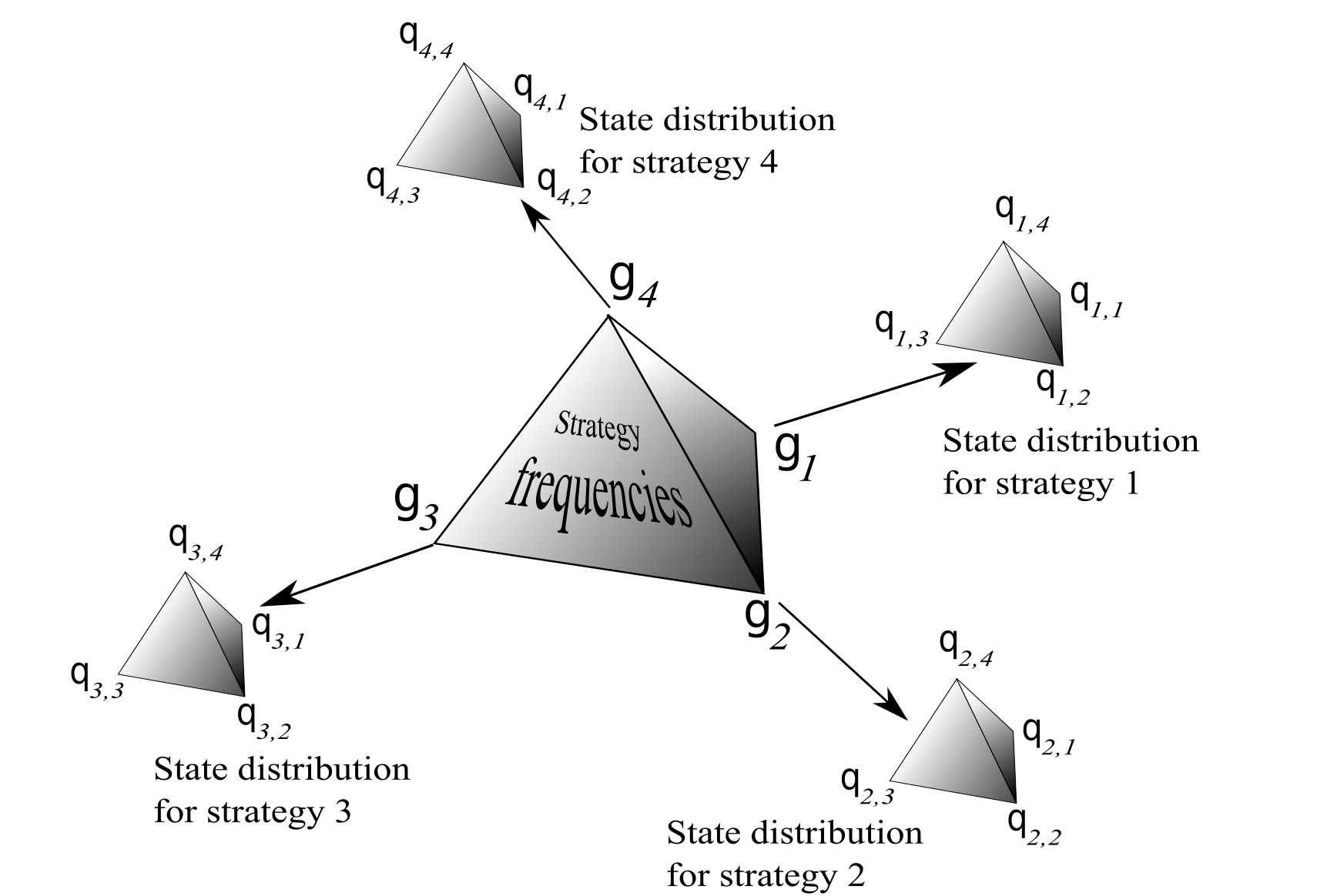}
\caption{Schematic presentation of the phase space of the system}
\end{figure}

\bigskip
We will use the multipopulation approach (\ref{multirep}) to replicator
dynamics (Argasinski 2006, 2012, 2013, 2018), where population can be
decomposed into subpopulations described by their own replicator dynamics.
Subsystems describing those subpopulations are completed by additional set
of replicator equations describing the dynamics of proportions of all
subpopulations. Then we can describe the distribution of states among $s$%
-strategists in related frequencies $q_{s}^{i}=n_{s}^{i}/\sum_{j}n_{s}^{j}$.
In effect, fertility payoff (\ref{fertility}) of the focal game can be
presented in the new indexing convention and in frequencies as:%
\begin{equation}
W_{s}^{i}(g,q)=\dfrac{\sum_{z}n_{s}^{z}O_{s}^{z}(g,q)\zeta _{s}^{z}(i)}{%
n_{s}^{i}}=\dfrac{\sum_{z}q_{s}^{z}O_{s}^{z}(g,q)\zeta _{s}^{z}(i)}{q_{s}^{i}%
}  \label{newfert}
\end{equation}%
Now we can derive the replicator dynamics, in a standard way, by rescaling
the growth equations $\dot{n}_{i}=n_{i}r_{i}=n_{i}\left( W_{i}-d_{i}\right) $
(where $r_{i}$ is the function describing growth rate) into frequency
equations (\ref{rep1}) which have form $\dot{q}_{i}=q_{i}\left(
r_{i}-\sum_{k}q_{k}r_{k}\right) =q_{i}\left( \left[ W_{i}-\sum_{k}q_{k}W_{k}%
\right] -\left[ d_{i}-\sum_{k}q_{k}d_{k}\right] \right) $, describing the
changes of distribution of the states for the $i$-th strategy, completed by
total population size equation $\dot{n}=n\sum_{i}q_{i}r_{i}$. Since%
\begin{equation}
\sum_{z\neq i}\dfrac{n_{s}^{z}}{n_{s}^{i}}c_{s}^{z,i}(g,q)=\sum_{z\neq i}%
\dfrac{q_{s}^{z}}{q_{s}^{i}}c_{s}^{z,i}(g,q)\ \ \ \ \text{and}\ \ \ \ \
\sum_{i}q_{s}^{i}\sum_{z\neq i}\dfrac{q_{s}^{z}}{q_{s}^{i}}%
c_{s}^{z,i}(g,q)=\sum_{i}\sum_{z\neq i}q_{s}^{z}c_{s}^{z,i}(g,q)
\label{qswitch}
\end{equation}
the respective bracketed term will be%
\begin{equation*}
q_{s}^{i}\left( \sum_{z\neq i}\dfrac{q_{s}^{z}}{q_{s}^{i}}%
c_{s}^{z,i}(g,q)-\sum_{k}\sum_{z\neq k}q_{s}^{z}c_{s}^{z,k}(g,q)\right) ,
\end{equation*}%
similarly for $\Lambda $ terms. Then we will obtain: \ \ \ 

\begin{eqnarray}
\dot{q}_{s}^{i} &=&{}q_{s}^{i}\left[ \left( W_{s}^{i}(g,q)-\bar{W}%
_{s}(g,q)\right) -\left( d_{s}^{i}(g,q)-\bar{d}_{s}(g,q)\right) +\left( \Phi
^{i}-\bar{\Phi}_{s}(q_{s})\right) \right.  \notag \\
&&\left. -\left( \Psi ^{i}-\bar{\Psi}_{s}(q_{s})\right) -\left(
c_{s}^{i}(g,q)-\bar{c}_{s}(g,q)\right) -\left( \Lambda ^{i}-\bar{\Lambda}%
(q_{s})\right) \right]  \notag \\
&&+\left( \sum_{z\neq
i}q_{s}^{z}c_{s}^{z,i}(g,q)-q_{s}^{i}\sum_{k}\sum_{z\neq
k}q_{s}^{z}c_{s}^{z,k}(g,q)\right)  \notag \\
&&+\left( \sum_{z\neq i}q_{s}^{z}\Lambda ^{z,i}-q_{s}^{i}\sum_{k}\sum_{z\neq
k}q_{s}^{z}\Lambda ^{z,k}\right) ,  \label{q}
\end{eqnarray}%
where 
\begin{eqnarray*}
\bar{W}_{s}(g,q) &=&\sum_{k}q_{s}^{k}W_{s}^{k}(g,q),\quad \bar{d}%
_{s}(g,q)=\sum_{k}q_{s}^{k}d_{s}^{k}(g,q), \\
\bar{\Phi}_{s}(q_{s}) &=&\sum_{k}q_{s}^{k}\Phi ^{k},\quad \bar{\Psi}%
_{s}(q_{s})=\sum_{k}q_{s}^{k}\Psi ^{k}, \\
\bar{c}_{s}(g,q) &=&\sum_{k}q_{s}^{k}c_{s}^{k}(g,q),\quad \text{and $\bar{%
\Lambda}(q_{s})=\sum_{k}q_{s}^{k}\Lambda ^{k}$}.
\end{eqnarray*}

\subsection{Case of two competing strategies}

In the special case where for all strategies we have only two states, the
above system reduces to the single equation. In addition it can be
simplified by application of well known form of the replicator dynamics for
two strategies $\dot{q}_{1}=q_{1}(1-q_{1})\left[ r_{1}-r_{2}\right] $. \ In
addition the terms describing the switching dynamics in (\ref{qswitch}) will
be also simplified and for both states will have forms%
\begin{eqnarray*}
\sum_{z\neq 1}\dfrac{n_{s}^{z}}{n_{s}^{1}}c_{s}^{z,1}(g,q) &=&\dfrac{\left(
1-q_{s}^{1}\right) }{q_{s}^{1}}c_{s}^{2,1}(g,q) \\
\sum_{z\neq 2}\dfrac{n_{s}^{z}}{n_{s}^{2}}c_{s}^{z,2}(g,q) &=&\dfrac{%
q_{s}^{1}}{\left( 1-q_{s}^{1}\right) }c_{s}^{1,2}(g,q),
\end{eqnarray*}%
leaving rates $c_{s}^{i}(g,q)$ reduce to%
\begin{equation*}
c_{s}^{1}(g,q)=c_{s}^{1,2}(g,q)\text{ \ \ \ and \ \ }%
c_{s}^{1}(g,q)=c_{s}^{1,2}(g,q).
\end{equation*}
since there is only single opposite state to switch. Then the external
bracketed term describing the impact of switching dynamics will be:

\begin{eqnarray*}
&&q_{s}^{1}\left( 1-q_{s}^{1}\right) \left( \left[ \dfrac{\left(
1-q_{s}^{1}\right) }{q_{s}^{1}}c_{s}^{2,1}(g,q)-c_{s}^{1,2}(g,q)\right] -%
\left[ \dfrac{q_{s}^{1}}{\left( 1-q_{s}^{1}\right) }%
c_{s}^{1,2}(g,q)-c_{s}^{2,1}(g,q)\right] \right) \\
&=&\left( 1-q_{s}^{1}\right) \left[ \left( 1-q_{s}^{1}\right)
c_{s}^{2,1}(g,q)-q_{s}^{1}c_{s}^{1,2}(g,q)\right] +q_{s}^{1}\left[ \left(
1-q_{s}^{1}\right) c_{s}^{2,1}(g,q)-q_{s}^{1}c_{s}^{1,2}(g,q)\right] \\
&=&\left( 1-q_{s}^{1}\right) c_{s}^{2,1}(g,q)-q_{s}^{1}c_{s}^{1,2}(g,q)
\end{eqnarray*}%

Similar form will have terms describing the background switching dynamics.
Then the equation describing the transitions between two states will be: 
\begin{align}
& \dot{q}_{s}^{1}=q_{s}^{1}\left( 1-q_{s}^{1}\right) \left[ \left(
W_{s}^{1}(g,q)-W_{s}^{2}(g,q)\right) -\left(
d_{s}^{1}(g,q)-d_{s}^{2}(g,q)\right) +\left( \Phi ^{1}-\Phi ^{2}\right)
-\left( \Psi ^{1}-\Psi ^{2}\right) \right]  \notag \\
& \,\,+\left[ \left( 1-q_{s}^{1}\right)
c_{s}^{2,1}(g,q)-q_{s}^{1}c_{s}^{1,2}(g,q)\right] +\left[ \left(
1-q_{s}^{1}\right) \Lambda ^{2,1}(g,q)-q_{s}^{1}\Lambda ^{1,2}(g,q)\right] ,
\label{2strategy}
\end{align}%
and the averaged values are not necessary.

\subsection{Selection of the strategies}

Now we can describe the selection of strategies by application of the
multipopulation approach \ref{multirep}). Then the above system \ ((\ref{q})
or (\ref{2strategy}))should be completed by the additional set of the
replicator equations describing the related frequencies of the other
strategies. Obviously the dynamics of state changes will not have direct
impact on the strategy frequencies (as well as on the population size) since
it will not change the number of strategy carriers. Then we have the
following system describing the selection:

\begin{equation}
\dot{g}_{s}={}g_{s}\left[ \left( \bar{W}_{s}(g,q)-\bar{W}(g,q)\right)
-\left( \bar{d}_{s}(g,q)-\bar{d}(g,q)\right) \right. \\
\left. +\left( \bar{\Phi}_{s}(q_{s})-\bar{\Phi}(g,q)\right) -\left( \bar{\Psi%
}_{s}(q_{s})-\bar{\Psi}(g,q)\right) \right] ,  \label{strat}
\end{equation}%
where%
\begin{eqnarray*}
\bar{W}(g,q) &=&\sum_{k}g_{k}\bar{W}_{k}(g,q),\quad \bar{d}%
(g,q)=\sum_{k}g_{k}\bar{d}_{k}(g,q), \\
\bar{\Phi}(g,q) &=&\sum_{k}g_{k}\bar{\Phi}_{k}(q_{k}),\quad \bar{\Psi}%
(g,q)=\sum_{k}g_{k}\bar{\Psi}_{k}(q_{k}).
\end{eqnarray*}
The above system should be completed by the equation on total population
size: 
\begin{equation}
\dot{n}=n\left[ \bar{W}(g,q)-\bar{d}(g,q)+\bar{\Phi}(g,q)-\bar{\Psi}(g,q)%
\right]  \label{popsize}
\end{equation}%
Note that for the uniform interaction rate $\tau _{F}$ for all strategies
the negative mortality bracketed terms $\left( \bar{d}_{s}(g,q)-\bar{d}%
(g,q)\right) $ and $\left( d_{s}^{1}(g,q)-d_{s}^{2}(g,q)\right) $ reduce to
the positive bracketed terms $\left( \bar{s}_{s}(g,q)-\bar{s}(g,q)\right) $
and $\left( s_{s}^{1}(g,q)-s_{s}^{2}(g,q)\right) $. this will not work for
nonuniform interaction rates. The above system can be easily extended to
density dependence by multiplying focal and background fertilities by some
juvenile recruitment mortality factor such as classical logistic suppression 
$(1-n/K)$.

\subsection{Obtained framework}

Summarizing the derivations from the previous subsections, we obtained the
following system containing switching dynamics, selection dynamics and the
population size:

\begin{eqnarray}
\dot{q}_{s}^{i} &=&{}q_{s}^{i}\left[ \left( W_{s}^{i}(g,q)-\bar{W}%
_{s}(g,q)\right) -\left( d_{s}^{i}(g,q)-\bar{d}_{s}(g,q)\right) +\left( \Phi
^{i}-\bar{\Phi}_{s}(q_{s})\right) \right.  \notag \\
&&\left. -\left( \Psi ^{i}-\bar{\Psi}_{s}(q_{s})\right) -\left(
c_{s}^{i}(g,q)-\bar{c}_{s}(g,q)\right) -\left( \Lambda ^{i}-\bar{\Lambda}%
(q_{s})\right) \right]  \notag \\
&&+\left( \sum_{z\neq
i}q_{s}^{z}c_{s}^{z,i}(g,q)-q_{s}^{i}\sum_{k}\sum_{z\neq
k}q_{s}^{z}c_{s}^{z,k}(g,q)\right)  \notag \\
&&+\left( \sum_{z\neq i}q_{s}^{z}\Lambda ^{z,i}-q_{s}^{i}\sum_{k}\sum_{z\neq
k}q_{s}^{z}\Lambda ^{z,k}\right) ,
\end{eqnarray}%
\begin{equation}
\dot{g}_{s}={}g_{s}\left[ \left( \bar{W}_{s}(g,q)-\bar{W}(g,q)\right)
-\left( \bar{d}_{s}(g,q)-\bar{d}(g,q)\right) \right. \\
\left. +\left( \bar{\Phi}_{s}(q_{s})-\bar{\Phi}(g,q)\right) -\left( \bar{\Psi%
}_{s}(q_{s})-\bar{\Psi}(g,q)\right) \right] ,
\end{equation}%
\begin{equation}
\dot{n}=n\left[ \bar{W}(g,q)-\bar{d}(g,q)+\bar{\Phi}(g,q)-\bar{\Psi}(g,q)%
\right]
\end{equation}

For the case of two states the switching dynamics reduces to%
\begin{align}
& \dot{q}_{s}^{1}=q_{s}^{1}\left( 1-q_{s}^{1}\right) \left[ \left(
W_{s}^{1}(g,q)-W_{s}^{2}(g,q)\right) -\left(
d_{s}^{1}(g,q)-d_{s}^{2}(g,q)\right) +\left( \Phi ^{1}-\Phi ^{2}\right)
-\left( \Psi ^{1}-\Psi ^{2}\right) \right]  \notag \\
& \,\,+\left[ \left( 1-q_{s}^{1}\right)
c_{s}^{2,1}(g,q)-q_{s}^{1}c_{s}^{1,2}(g,q)\right] +\left[ \left(
1-q_{s}^{1}\right) \Lambda ^{2,1}(g,q)-q_{s}^{1}\Lambda ^{1,2}(g,q)\right] ,
\end{align}%
Depending on the causal structure underlying the modelled phenomenon, payoff
functions $W_{s}^{i}$, $d_{s}^{i}$ and $c_{s}^{i}$ can be replaced by more
complex trade-off functions $V_{s}^{i}$, $Z_{s}^{i,j}$, $X_{s}^{i}$ or $%
Y_{s}^{i}$. For different interaction rates for different states/roles we
can use the functions (\ref{deathrate1},\ref{deathrate2}) and (\ref{tau1F},%
\ref{tau2F}). Then the focal game mortality payoffs $d_{s}^{i}(g,q)$ should
be replaced by general death rates 
\begin{equation}
D_{i}^{1}(g,q)=d_{i}^{1}(g,q)=1-s_{i}^{1}(g,q)\text{ \ \ for \ \ role \ \ 1
(in minority)}
\end{equation}%
\begin{equation}
D_{i}^{2}(g,q)=d_{i}^{2}(g,q)\dfrac{\sum_{i}g_{i}q_{i}^{1}}{%
\sum_{i}g_{i}(1-q_{i}^{1})}=\left( 1-s_{i}^{2}(g,q)\right) \dfrac{%
\sum_{i}g_{i}q_{i}^{1}}{\sum_{i}g_{i}(1-q_{i}^{1})}\text{ \ \ for role 2 (in
majority),}
\end{equation}

or by respective alternative form in the opposite situation. However, we
will see in the later sections that we can successfully build a model
limited to the case when one particular role is always in minority.

\subsection{Separation of the timescales between demographic and switching
dynamics}

Now let us examine assumption that demographic events are separate from
switching events. Let us get back to the equation (\ref{basicmalth}):%
\begin{eqnarray}
\dot{n}_{s,i} &=&n_{s,i}\tau _{F}\left[ \left(
W_{s,i}^{F}(g,q)-d_{s,i}^{F}(g,q)\right) +\left( \sum_{z\neq i}\dfrac{n_{s,z}%
}{n_{s,i}}c_{s,z,i}^{F}(g,q)-c_{s,i}^{F}(g,q)\right) \right]  \notag \\
&&+n_{s,i}\tau _{B}\left[ \left( W_{B,i}-m_{B,i}\right) +\left( \sum_{z\neq
i}\dfrac{n_{s,z}}{n_{s,i}}c_{B,z,i}-c_{B,i}\right) \right] .
\end{eqnarray}%
Similarly to the separation of the focal game from the background events, we
can reindex the rates to distinguish between those two classes of events.
Then demographic events will occur with rate $\tau _{dem}^{j}$ while state
changing events will occur at rates $\tau _{st}^{i}$ Then demographic event
will occur at the intensity $\tau _{dem}=\sum_{l}\tau _{dem}^{l}$ and when
it occurs then it will be $j$-th type event with probability $p_{dem}^{j}=%
\dfrac{\tau _{dem}^{j}}{\sum_{l}\tau _{dem}^{l}}$. Similarly for state
switching events we have $\tau _{st}=\sum_{l}\tau _{st}^{l}$ and $p_{st}^{j}=%
\dfrac{\tau _{st}^{j}}{\sum_{l}\tau _{st}^{l}}$. We can extract focal
demographic and switching events 
\begin{eqnarray}
\dot{n}_{s,i} &=&n_{s,i}\tau _{F}\left[ \tau _{dem}\left(
W_{s,i}^{F}(g,q)-d_{s,i}^{F}(g,q)\right) +\tau _{st}\left( \sum_{z\neq i}%
\dfrac{n_{s,z}}{n_{s,i}}c_{s,z,i}^{F}(g,q)-c_{s,i}^{F}(g,q)\right) \right] 
\notag \\
&&+n_{s,i}\tau _{B}\left[ \tau _{dem}\left( W_{B,i}-m_{B,i}\right) +\tau
_{st}\left( \sum_{z\neq i}\dfrac{n_{s,z}}{n_{s,i}}c_{B,z,i}-c_{B,i}\right) %
\right] .
\end{eqnarray}%
Then we can assume that $\tau _{dem}<<\tau _{st}$ we can separate the
timescales. In the resulting $q$ equations $\tau _{st}$ can be extracted
from the bracket and set to 1 by timescale adjustment, in effect $\tau
_{dem} $ will be replaced by $\phi =\dfrac{\tau _{dem}}{\tau _{st}}$. In the
limit $\phi \rightarrow 0$ we obtain fast system describing the state
changes: 
\begin{eqnarray*}
\dot{q}_{s}^{i} &=&{}\left(
\sum_{z}q_{s}^{z}c_{s}^{z,i}(g,q)-q_{s}^{i}\sum_{k}%
\sum_{z}q_{s}^{z}c_{s}^{z,k}(g,q)\right) \\
&&+\left( \sum_{z}q_{s}^{z}\Lambda
^{z,i}-q_{s}^{i}\sum_{k}\sum_{z}q_{s}^{z}\Lambda ^{z,k}\right) \\
&&-q_{s}^{i}\left[ \left( c_{s}^{i}(g,q)-\bar{c}_{s}(g,q)\right) -\left(
\Lambda ^{i}-\bar{\Lambda}(q_{s})\right) \right]
\end{eqnarray*}%
and the slow selection system driven by the equilibria of the above state
changing system which will have the same form as (\ref{strat}) completed by
the equation on the population size (\ref{popsize}). We can imagine the
opposite situation when state changes are much slower than the demographic
dynamics, as for example in the process of the senescence (then we should
limit to the mortality payoffs of the focal game since newborns will also
follow slow ageing process, thus they cannot rapidly mature and enter the
game). Then we can assume that $\tau _{st}<<\tau _{dem}$ and in the similar
way obtain the model where selection dynamics is the fast system while the
dynamics of state changes is slow, since fast demographic dynamics will
rapidly set the bracketed terms in the first line of equation (\ref{q}) to
zero. Note that above approaches will be applicable only in the cases where
equilibria of fast systems exist.

\section{ Special case: Stage structured population\protect\bigskip}

Now let us consider the case of the population where individuals are
affected by some irreversible process such as developmental cycle or
senescence which can be interpreted as the accumulation of damages.\ The
phenomenon of this kind can be also considered as the state changing
process. Then we should modify our approach to describe the stepwise
incremental process with $x$ possible levels. Then $c_{s}^{i,i+1}$ and $%
\Lambda ^{i,i+1}$ are intensities of the next step in the process caused by
focal type of interaction or by background dynamics. Then the term%
\begin{equation*}
\sum_{z\neq i}\dfrac{n_{s}^{z}}{n_{s}^{i}}c_{s}^{z,i}(g,q)-c_{s}^{i}(g,q)
\end{equation*}%
in equation (\ref{malth}) reduces to%
\begin{equation*}
\dfrac{n_{s}^{i-1}}{n_{s}^{i}}c_{s}^{i-1,i}(g,q)-c_{s}^{i,i+1}(g,q)
\end{equation*}%
(the term describing the background state changes will have similar form).
We assume that all newborns are in stage 0 and there are no interactions
between them, thus all of them will pass to the stage 1 with intensity $%
\Lambda ^{0,1}$ and in the last stage the individuals will only die with
intensity $\Psi ^{x}$. Thus the aggregated fertility of the strategy $s$
will be%
\begin{equation*}
\tilde{W}_{s}(g,q)=\sum_{j}n_{s}^{j}W_{s}^{j}(g,q),
\end{equation*}%
and to obtain per capita value it should be divided by respective number of
0 stage individuals (similar function will be for the background fertilities 
$\Phi $). We should also describe the causal chain of the focal interaction,
since only survivors of the interactions should change the state, post
mortality switching function $Z_{i}^{j,x}$ (\ref{m-s}) should be applied and
replace the function $c_{i}^{j,x}$ (application of the post-switching
survival described by function $X_{i}^{j}$ (\ref{s-s})\ will lead to
immortality with $c_{i}^{l}$ converging to 1). leading to the following form
of (\ref{malth}): 
\begin{eqnarray*}
\dot{n}_{s}^{0} &=&{}n_{s}^{0}\left[ \frac{\tilde{W}_{s}(g,q)}{n_{s}^{0}}%
-\Lambda ^{0,1}\right] \\
\dot{n}_{s}^{i} &=&{}n_{s}^{i}\left[ \left( \dfrac{n_{s}^{i-1}}{n_{s}^{i}}%
Z_{s}^{i-1,i}(g,q)-Z_{s}^{i,i+1}(g,q)\right) -(1-s_{s}^{i}(g,q))\right] \\
&&+n_{s}^{i}\left( \dfrac{n_{s}^{i-1}}{n_{s}^{i}}\Lambda ^{i-1,i}-\Lambda
^{i,i+1}\right) -n_{s}^{i}\Psi ^{i} \\
\dot{n}_{s}^{x} &=&{}n_{s}^{x}\left[ \dfrac{n_{s}^{x-1}}{n_{s}^{x}}\left(
Z_{s}^{i-1,i}(g,q)+\Lambda _{s}^{x-1,x}(g,q)\right) -\Psi ^{x}\right] .
\end{eqnarray*}%
The above system can be presented as 
\begin{align}
\dot{n}_{s}^{0}={}& \tilde{W}_{s}(g,q)-n_{s}^{0}\Lambda ^{0,1}
\label{stage1} \\
\dot{n}_{s}^{i}={}& n_{s}^{i-1}Z_{s}^{i-1,i}(g,q)-n_{s}^{i}\left(
Z_{s}^{i,i+1}(g,q)-d_{s}^{i}(g,q)\right)  \label{stage2} \\
& +n_{s}^{i-1}\Lambda ^{i-1,i}-n_{s}^{i}\left( \Lambda ^{i,i+1}-\Psi
^{i}\right) .  \label{stage3} \\
\dot{n}_{s}^{x}={}& n_{s}^{x}\left[ n_{s}^{x-1}\Lambda
_{s}^{x-1,x}(g,q)-n_{s}^{x}\Psi _{x}\right] .  \label{stage4}
\end{align}%
Let us derive the replicator dynamics describing the state switching
dynamics. Note that for average switching payoffs we have that 
\begin{equation*}
\sum_{i}q_{s}^{i}\left( \dfrac{q_{s}^{i-1}}{q_{s}^{i}}%
Z_{s}^{i-1,i}(g,q)-Z_{s}^{i,i+1}(g,q)\right) \\
=\sum_{i}q_{s}^{i-1}Z_{s}^{i-1,i}(g,q)-\sum_{i}q_{s}^{i}Z_{s}^{i,i+1}(g,q)=0,
\end{equation*}%
since by definition in the last state class $x$ we have $%
Z_{s}^{x,x+1}(g,q)=0 $ (similarly for $\Lambda $ terms). This simplifies the
bracketed terms describing switching payoffs excess from the average value.
We will derive replicator dynamics describing the frequencies of stages from 
$1$ to $x$, thus fertility payoffs in those stages are $0$ by definition and
then average fertility for carrier subpopulation is%
\begin{equation*}
\bar{W}_{s}(g,q)=\sum_{j}q_{s}^{j}W_{s}^{j}(g,q)=\frac{\tilde{W}_{s}(g,q)}{%
n_{s}}.
\end{equation*}%
since in this case $W_{s}^{j}(g,q)$ (\ref{newfert}) describes per capita
growth of newborn subpopulation. Similar situation is for background
switching subsystem. Therefore system (\ref{q}) will be reduced to: 
\begin{eqnarray}
\dot{q}_{s}^{i} &=&q_{s}^{i}\left[ \left( s_{s}^{i}(g,q)-\bar{s}%
_{s}(g,q)\right) -\left( \Psi ^{i}-\Psi _{s}(q_{s})\right) -\bar{W}_{s}(g,q)-%
\bar{\Phi}_{s}(q_{s})\right.  \notag \\
&&\left. -Z_{s}^{i,i+1}(g,q)-\Lambda ^{i,i+1}\right] +q_{s}^{i-1}\left[
Z_{s}^{i-1,i}(g,q)+\Lambda ^{i-1,i}\right]  \label{qstage}
\end{eqnarray}%
Equations describing the selection of the strategies and the population size
will be

\begin{eqnarray*}
\dot{g}_{s} &=&{}g_{s}\left[ \left( \bar{W}_{s}(g,q)-\bar{W}(g,q)\right)
+\left( \bar{s}_{s}(g,q)-\bar{s}(g,q)\right) \right. \\
&&\left. {}+\left( \bar{\Phi}_{s}(q_{s})-\bar{\Phi}(g,q)\right) -\left( \bar{%
\Psi}_{s}(q_{s})-\bar{\Psi}(g,q)\right) \right] \\
\dot{n} &=&{}n\left[ \bar{W}(g,q)+\bar{s}(g,q)-1+\bar{\Phi}(g,q)-\bar{\Psi}%
(g,q)\right] .
\end{eqnarray*}%
Note that (\ref{qstage}) is attracted by the surface 
\begin{equation*}
q_{s}^{i}=q_{s}^{i-1}\frac{Z_{s}^{i-1,i}(g,q)+\Lambda ^{i-1,i}}{D},
\end{equation*}%
where 
\begin{equation*}
D=\left( s_{s}^{i}(g,q)-\bar{s}_{s}(g,q)\right) -\left( \Psi ^{i}-\Psi
_{s}(q_{s})\right) -\bar{W}_{s}(g,q)-\bar{\Phi}%
_{s}(q_{s})-Z_{s}^{i,i+1}(g,q)-\Lambda ^{i,i+1}.
\end{equation*}

\subsection{Limit case: classical age structured models of life history
evolution}

Interesting is the case when all individuals in the population (not only
those involved in the interactions) are subject of switching dynamics.
Assume that the switching probabilities are equal to $c$ for all states and
strategies and there is no background switching dynamics $\Lambda $ (with
exception of the state 0 where there are no focal game switching payoffs, we
assume $\Lambda ^{0,1}=1$) and background fertility and mortality payoffs $%
\Phi $ and $\Psi $. This case will describe the phenomenological aging or
senescence. The simple discrete system presented here is insufficient for
description of the detailed game dynamics where payoff functions depend on
the population composition $(q,g)$. However, it can be used for description
of the simpler case when payoffs of different strategies depend only on
their age or on some physiologic strategy of allocation of resources (such
as tradeoff in investment in reproduction or in somatic\ repair). Therefore
demographic payoffs will be functions of\ the value $p^{j}$ of some
physiological trait in age class $j$ ($W^{j}(p_{s}^{j})$ and $%
s^{j}(p_{s}^{j})$). Then vector $p_{s}$ is the physiological life history
strategy which replaces the argument $g$ describing strategy frequencies.\
This can be used for modelling of the selection of life history strategies
(Stearns 1992, Roff 2002). Then post-mortality switching function will be
simply%
\begin{equation}
Z_{i}^{j,j+1}(p_{s}^{j})=cs_{i}^{j}(p_{s}^{j}),  \label{zs}
\end{equation}%
therefore for $c=1$ survival function $s$ replaces switching function $Z$
and the system (\ref{stage1}-\ref{stage4}) of equations becomes:%
\begin{eqnarray}
\dot{n}_{s}^{0} &=&n_{s}^{0}\bar{W}_{s}(p_{s},q_{s})-n_{s}^{0}  \label{les1}
\\
\dot{n}_{s}^{i} &=&n_{s}^{i-1}s^{i-1,i}(p_{s}^{i-1})-n_{s}^{i}.  \label{les2}
\end{eqnarray}

For single strategy $s$ this will be the continuous time equivalent of the
Leslie matrix model if the above model will reflect the actual age described
by delays between censuses of the population. If the duration of the age
classes will be $\gamma $, then the system (\ref{les1},\ref{les2}) will be:

\begin{eqnarray}
\dot{n}_{s}^{0}(t) &=&n_{s}^{0}(t)\bar{W}_{s}(p_{s},q_{s})-n_{s}^{0}(t) \\
\dot{n}_{s}^{i}(t) &=&n_{s}^{i-1}(t-\gamma
)s^{i-1,i}(p_{s}^{i-1})-n_{s}^{i}(t),
\end{eqnarray}%
where $s_{s}^{i-1,i}$ can be interpreted as the aggregated exponential
survival between censuses and $\bar{W}_{s}(p_{s},q_{s})=%
\sum_{j}q_{s}^{j}W^{j}(p_{s}^{j})$. Above system can be rescaled to the
dynamics describing the age structure and completed by replicator dynamics
describing the selection of the strategies and the equation for population
size. In effect in (\ref{qstage}) $\Psi $, $\bar{\Phi}_{s}(q_{s})$ and $%
\Lambda $ are not present, as we assumed above. In addition (\ref{zs})
implies that $Z_{s}^{i,i+1}$ is replaced by $s_{s}^{i}$ which cancels out
and $s_{i}^{j-1}$replaces $Z_{s}^{i-1,i}$. In effect we obtain
system:\bigskip

\begin{eqnarray}
\dot{q}_{s}^{i}(t) &=&q_{s}^{i-1}(t-\gamma
)s_{s}^{i-1}(p_{s}^{i-1})-q_{s}^{i}(t)\left[ \bar{s}_{s}(p_{s},q_{s})+\bar{W}%
_{s}(g)\right] \\
\dot{g}_{s} &=&g_{s}\left[ \left( \bar{W}_{s}(p_{s},q_{s})-\bar{W}(g)\right)
+\left( \bar{s}_{s}(p_{s},q_{s})-\bar{s}(g)\right) \right] \\
\dot{n} &=&n\left[ \bar{W}(g)+\bar{s}(g)-1\right] ,
\end{eqnarray}%
where $\bar{s}_{s}(p_{s},q_{s})=\sum_{j}q_{s}^{j}s^{j}(p_{s}^{j})$, $\bar{W}%
(g)=\sum_{s}\bar{W}_{s}(p_{s},q_{s})$ and $\bar{s}(g)=\sum_{s}\bar{s}%
_{s}(p_{s},q_{s})$. The more complex models allowing for description of
frequency dependent game dynamics where payoffs depend on the vector of the
strategy frequencies $g$, will be subject of the next paper. The main
problem is that in this case life cycle of the individual cannot be
discretized since the mortality between censuses is shaped by trajectory of
the population composition.

\section{Example: Owner-Intruder game with explicit role distribution and
the underlying dynamics.\protect\bigskip}

Owner-Intruder game model was introduced by John Maynard Smith (1982) as a
simple bimatrix game model of conflict for property, where individuals may
randomly act in both roles of Owner and Intruder with equal probability. It
is still analyzed (Cressman and K\v{r}ivan 2019) since it is one of the
basic examples of asymmetric matrix games. Model suggests that the strategy
called Bourgeois, defending the property while respecting the property of
others, by playing Dove as Intruder, should be favored by natural selection.
Later works suggest that the opposite strategy called Anti-bourgeois or
Vagabond (Grafen 1987, Eschel and Sansone 1995) can be also justified in
some cases. The problem still needs the general explanation (Sherratt and
Mesterton-Gibbons 2015). The classic model, as every bimatrix game ignores
the proportion between both subgroups, which may lead to false predictions
of the models (Argasinski 2006). In addition it ignores the dynamics of role
changes. This aspect was explicitly considered in Kokko et al. (2006) and
later in K\v{r}ivan et al. (2018). Alternatively, the time spent in each
role can be considered (Hinsch and Komdeur 2010). Our model will be
extension of the demographic formulation of Hawk-Dove game (Argasinski and
Broom 2013,2017,2018), based on similar assumptions to Kokko et al. 2006 and
K\v{r}ivan et al. (2018, but for simplicity we will not include the time
constraints), but it will be more focused on the dynamics and the
explanation of underlying mechanisms, since the previous papers were more
focused on equilibria and the static analysis of the outcomes of the
strategy selection. Due to our focus on the dynamics and the processes we
will ignore asymmetries between roles (Leimar and Enquist 1984, Korona 1989,
1991) for simplicity of payoff functions. So let us start the development of
the new model.

\begin{figure}[h!]
\centering
\includegraphics[width=14cm]{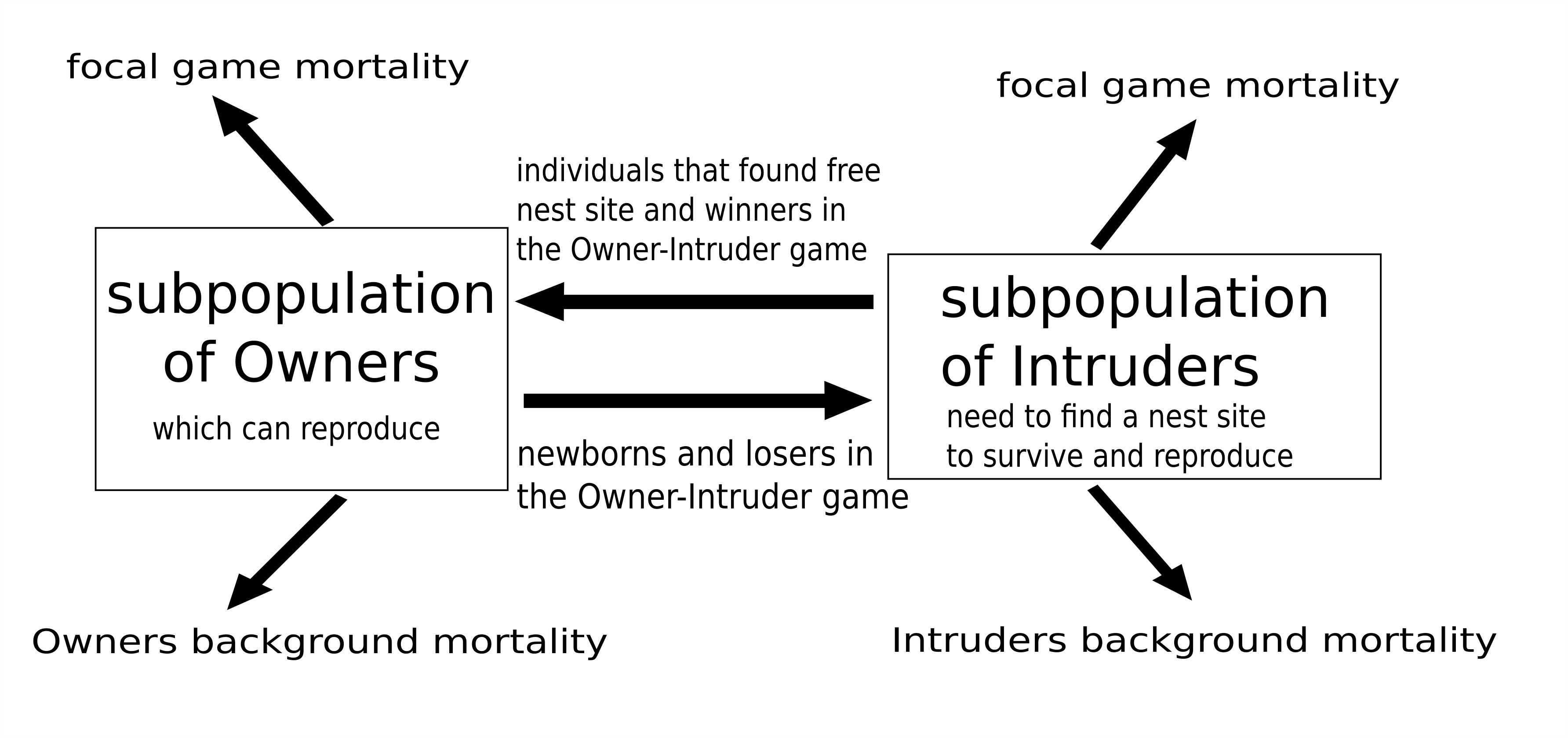}
\caption{Structure of the model. Diagram of fluxes of indiviuals between Owner and Intruder roles and the fluxes caused by different mortality soruces.}
\end{figure}

We have two states, Owner of the Habitat and the Intruder. This leads to the
two subpopulations, which strategic compositions are affected by respective
mortalities and fertilities and fluxes between both roles (see Fig.2 for
detailed presentation). Only Owners can reproduce and this factor will be
described by their background fertility $\Phi $ and produced newborns will
become homeless intruders. Intruders will randomly check the nest sites at
the constant inspection rate, settle down when the nest site is free or play
with the Owner. Fight for the habitat, a round of Hawk-Dove game played by
Owner and Intruder, will be associated with the risk of death described by
survival payoffs. The difference between basic Hawk-Dove game (Argasinski
and Broom 2013,2017,2018) and a new model is that in the new model opponents
are randomly drawn from separate pools (subpopulations of Owners and
Intruders). This leads to different numbers of interactions per time unit
(described by different interaction rates) for opposite roles (see Fig 3).
By application of the concept of interaction rates (Argasinski and Broom
2017), this situation can be modelled by (\ref{deathrate1},\ref{deathrate2})
and (\ref{tau1F},\ref{tau2F}).

\begin{figure}[h!]
\centering
\includegraphics[width=14cm]{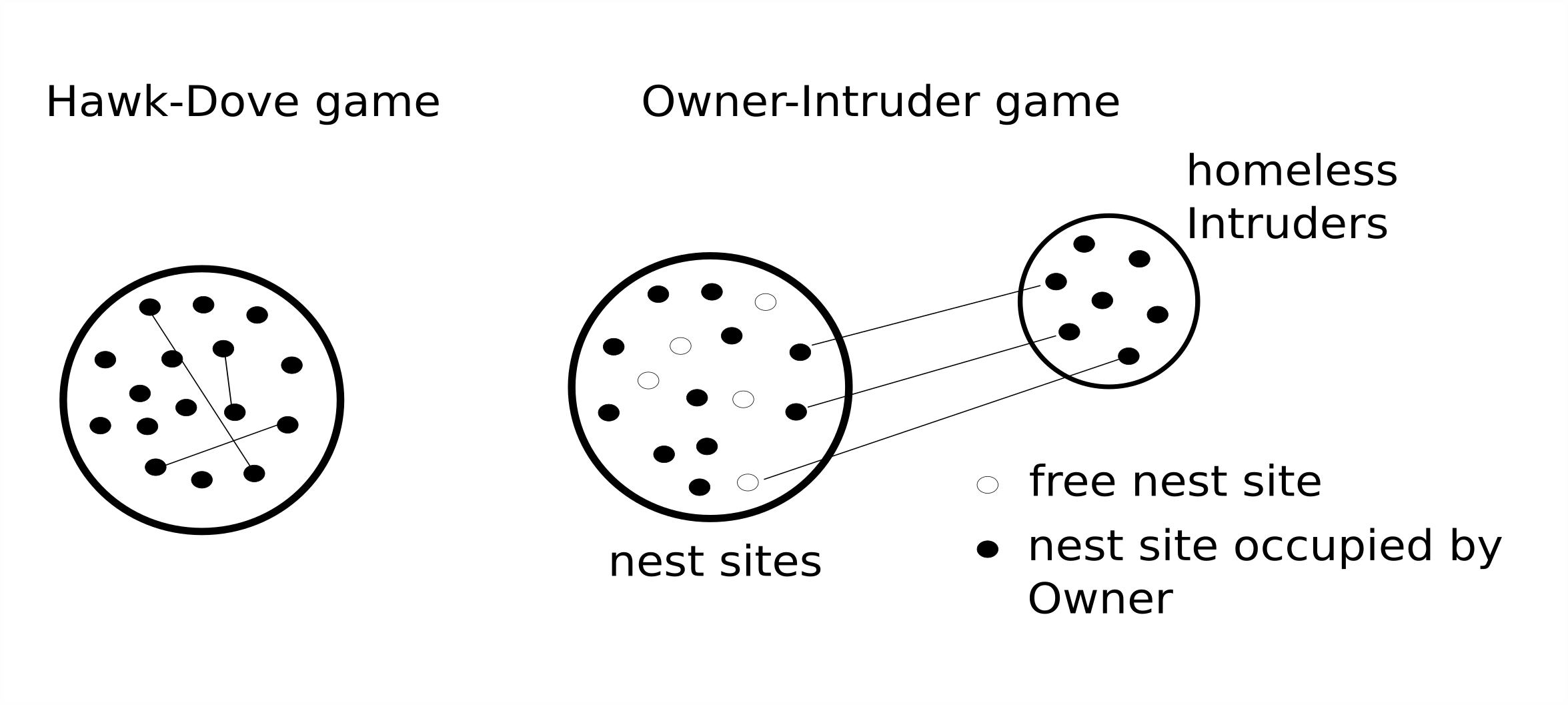}
\caption{In Hawk-Dove game we have panmictic population with a random pair
formation, thus all indiviuals will interact at the same rate. In the
Owner-Intruder game Intruders check the nest-sites at the constant
inspection rate. Then, for example, if we have 16 nest sites inhabited by 11
Owners and surrounded by 7\ Intruders, as depicted on the figure, for three
inspections per capita probability of fight for Owner is 2/11, while for
Intruder 2/7. Therefore, the difference between Hawk-Dove game and
Owner-Intruder game mainly depends on different interaction rates for
different roles.}
\end{figure}

Behavioral patterns (actions) are Hawk (aggressive) and Dove (peaceful).
Individual strategy is defined as the pair of behavioral patterns associated
to the state. Thus, following the classic Maynard-Smith terminology (1982)
we have Pure Hawk, Pure Dove, Bourgeois (Hawk when Owner, Dove when
Intruder) and Anti-Bourgeois (opposite to Bourgeois). For each strategy the
distribution of roles is described by $q_{i}^{O}=1-q_{i}^{I}$ Now we should
define payoffs. Action-specific survival payoffs will be described by matrix:%
$\bigskip $

$\ \ \ \ \ \ \ \ \ 
\begin{array}{cc}
H & D%
\end{array}%
$\newline
$S=%
\begin{array}{c}
H \\ 
D%
\end{array}%
\left[ 
\begin{array}{cc}
s & 1 \\ 
1 & 1%
\end{array}%
\right] $\bigskip \newline
and will be the same for both states. The switch payoffs will be \bigskip

$\qquad \ \ \ \ \ \ 
\begin{array}{cc}
H & D%
\end{array}%
$\newline
$C^{O}=%
\begin{array}{c}
H \\ 
D%
\end{array}%
\left[ 
\begin{array}{cc}
0.5 & 0 \\ 
1 & 0.5%
\end{array}%
\right] $ \ \ \ \ for Owners\bigskip

$\qquad \ \ \ \ \ \ \ 
\begin{array}{cc}
H & D%
\end{array}%
$\newline
$C^{I}=%
\begin{array}{c}
H \\ 
D%
\end{array}%
\left[ 
\begin{array}{cc}
0.5 & 1 \\ 
0 & 0.5%
\end{array}%
\right] $ \ \ \ \ for Intruders\bigskip \newline
Since only survivors can switch we should introduce the mortality-switching
trade-off functions $Z$ (\ref{m-s}). Thus for simplicity we should describe
the model in terms of survival $s_{s}^{i}(g,q)=1-d_{s}^{i}(g,q)$. We should
start the derivation of functions $Z^{O}$ and$\ Z^{I}$ by multiplying
entries of $S$ and $C^{O}$ ($S$ and $C^{I}$) elementwise (the only
difference will be $0.5s$ for Hawk-Hawk interaction), leading to:

$\qquad \ \ \ \ \ \ \ \ 
\begin{array}{cc}
H & D%
\end{array}%
$\newline
$Z^{O}=%
\begin{array}{c}
H \\ 
D%
\end{array}%
\left[ 
\begin{array}{cc}
0.5s & 0 \\ 
1 & 0.5%
\end{array}%
\right] $ \ \ \ \ for Owners\bigskip

$\qquad \ \ \ \ \ \ \ 
\begin{array}{cc}
H & D%
\end{array}%
$\newline
$Z^{I}=%
\begin{array}{c}
H \\ 
D%
\end{array}%
\left[ 
\begin{array}{cc}
0.5s & 1 \\ 
0 & 0.5%
\end{array}%
\right] $ \ \ \ \ for Intruders\bigskip \newline
Now we should derive the arguments of those payoff functions which are
distributions of actions (more specifically proportion of Hawk playing
strategies among individuals of both states:%
\begin{equation}
H^{O}=\dfrac{\sum_{H}g_{i}q_{i}^{O}}{\sum_{i}g_{i}q_{i}^{O}}\text{ \ \ and \
\ \ }H^{I}=\dfrac{\sum_{H}g_{i}q_{i}^{I}}{\sum_{i}g_{i}q_{i}^{I}}=\dfrac{%
\sum_{H}g_{i}\left( 1-q_{i}^{O}\right) }{\sum_{i}g_{i}\left(
1-q_{i}^{O}\right) }  \label{HoHi}
\end{equation}%
(and $\sum_{H}g_{i}q_{i}^{O}$ ($\sum_{H}g_{i}q_{i}^{I}$)means sum over
strategies playing Hawk as Owners (Intruders)). Proportions of Owners and
Intruders will be $\sum_{i}g_{i}q_{i}^{O}$ and $1-\sum_{i}g_{i}q_{i}^{O}$.
Individuals in the population compete for $K$ available nest sites (Hui
2006), then $K-n\sum_{i}g_{i}q_{i}^{O}$ will be number of free nest sites in
the population. Then the Intruder will check the randomly chosen nest site
and play the game with the Owner with probability%
\begin{equation*}
p_{g}=\frac{n\sum_{i}g_{i}q_{i}^{O}}{K},
\end{equation*}%
or stay there when it is free with probability%
\begin{equation*}
1-p_{g}=\frac{K-n\sum_{i}g_{i}q_{i}^{O}}{K}.
\end{equation*}%
Then the respective action specific survival payoffs per nest site
inspection will be:

\begin{eqnarray}
s_{H}^{O} &=&H^{I}s+(1-H^{I})=1-H^{I}\left( 1-s\right) ,  \label{survOH} \\
s_{H}^{I} &=&\left( 1-p_{g}\right) +p_{g}\left[ H^{O}s+(1-H^{O})\right] 
\notag \\
&=&1-H^{O}\left( 1-s\right) \frac{n\sum_{i}g_{i}q_{i}^{O}}{K},
\label{survIH} \\
s_{D}^{O} &=&s_{D}^{I}=1.  \label{survD}
\end{eqnarray}%
We have that $s_{H}^{O}$ and $s_{H}^{I}$ are always smaller than $s_{D}^{O}$
and $s_{D}^{I}$. Switching payoffs per nest site inspection will be:

\begin{eqnarray}
Z_{H}^{O,I} &=&H^{I}0.5s+(1-H^{I})0=H^{I}0.5s  \label{swOIH} \\
Z_{D}^{O,I} &=&H^{I}+(1-H^{I})0.5=0.5(1+H^{I})  \label{swOID} \\
Z_{H}^{,I,O} &=&\left( 1-p_{g}\right) +p_{g}\left[ H^{O}0.5s+(1-H^{O})\right]
\notag \\
&=&1-H^{O}\left[ 1-0.5s\right] \frac{n\sum_{i}g_{i}q_{i}^{O}}{K}
\label{swIOH} \\
Z_{D}^{,I,O} &=&\left( 1-p_{g}\right) +p_{g}0.5(1-H^{O})=  \notag \\
&=&1-\frac{n\sum_{i}g_{i}q_{i}^{O}}{K}0.5(1+H^{O})=  \notag \\
&=&1-0.5\left( \sum_{i}g_{i}q_{i}^{O}+\sum_{H}g_{i}q_{i}^{O}\right) \frac{n}{%
K}  \label{swIOD}
\end{eqnarray}%
and always $Z_{H}^{O,I}\leq Z_{D}^{O,I}$ and $Z_{D}^{,I,O}\leq Z_{H}^{,I,O}$%
. In this case there are no fertility payoffs for strategies. Only Owners
reproduce at the rate $\Phi $ for all strategies and their fertility feeds
the Invaders subpopulation. However the reproduction is not related to the
focal interaction, thus it can be regarded as background fertility $\Phi $.
Then the fertility for the strategy, describing per capita growth of
invaders will be 
\begin{equation}
\Phi _{s}^{I}=\frac{q_{s}^{O}}{1-q_{s}^{O}}\Phi  \label{phiI}
\end{equation}%
We also assume the background mortalities for both states $\Psi _{O}<\Psi
_{I}$ since Owners are safer. In addition there is no background switching,
since individuals cannot leave the nest site without a reason. Above payoff
functions describe survival and switching outcomes of the average
interaction which is fight for the occupied nest site. However in this case,
we will have different interaction rates for different states, similarly to (%
\ref{tau1F},\ref{tau2F}), which means that Owners will play different number
of game rounds per single time unit than Intruders (see Fig.3). Owners play
the game with Intruders checking the nest sites and the probability that the
nest site will be invaded is $\frac{n\left( 1-\sum_{i}g_{i}q_{i}^{O}\right) 
}{K}$ when the number of Intruders is not greater than $K$. Otherwise, $%
\frac{K}{n\left( 1-\sum_{i}g_{i}q_{i}^{O}\right) }$ is the probability that
the Intruder will check single nest site, when the number of intruders
exceeds $K$. Therefore the ratio of interaction rates is%
\begin{equation}
\dfrac{\tau _{F}^{O}}{\tau _{F}^{I}}=\dfrac{n\left(
1-\sum_{i}g_{i}q_{i}^{O}\right) }{K}  \label{tauratio}
\end{equation}%
and it should be multiplied by Owners payoffs to obtain death and switching
rates similarly to (\ref{deathrate1},\ref{deathrate2}). Therefore, we can
assume that the timescale of our model is adjusted to the constant Intruders
inspection rate of the nest sites, however for simplicity we will limit
ourselves to the cases when $n\left( 1-\sum_{i}g_{i}q_{i}^{O}\right) <K$
i.e. the number of Intruders is smaller than total number of nest sites.
This implies constant inspection rate for Intruders.

\subsection{Switching dynamics}

We can use (\ref{2strategy}), which is%
\begin{eqnarray}
\dot{q}_{s}^{x} &=&{}q_{s}^{x}\left( 1-q_{s}^{x}\right) \left[ \left( \Phi
_{x}-\Phi _{y}\right) -\left( \Psi _{x}-\Psi _{y}\right) -\left(
D_{s}^{x}(g,q)-D_{s}^{y}(g,q)\right) \right]  \notag \\
&&{}+\left[ \left( 1-q_{s}^{x}\right)
Z_{s}^{y,x}(g,q)-q_{s}^{x}Z_{s}^{x,y}(g,q)\right] ,  \label{swgeneral}
\end{eqnarray}%
to describe the switching dynamics for different strategies. We can use
survival payoffs (\ref{survOH}),(\ref{survIH}) and (\ref{survD}) to (\ref%
{deathrate1}) and (\ref{deathrate2}) for derivation of mortality bracket
(which will be positive in equation for fraction of Owners) 
\begin{equation*}
\left( D_{s}^{I}(g,q)-D_{s}^{O}(g,q)\right) =\left( \left[ 1-s_{s}^{I}(g,q)%
\right] -\left[ 1-s_{s}^{O}(g,q)\right] \frac{n\left(
1-\sum_{i}g_{i}q_{i}^{O}\right) }{K}\right) ,
\end{equation*}%
similarly we will use switching payoffs (\ref{swOIH}), (\ref{swOID}), (\ref%
{swIOH}) and (\ref{swIOD}) for derivation of the bracketed term $\left[
\left( 1-q_{s}^{x}\right) Z_{s}^{y,x}(g,q)-q_{s}^{x}Z_{s}^{x,y}(g,q)\right] $%
. Then, since $\Phi _{s}^{O}=0$ in our case and due to (\ref{phiI}), the
bracket ${}q_{s}^{x}\left( 1-q_{s}^{x}\right) \left( \Phi _{x}-\Phi
_{y}\right) $ reduces to negative term $\left( q_{s}^{O}\right) ^{2}\Phi $.
Then (\ref{swgeneral}) will be

\begin{eqnarray*}
\dot{q}_{s}^{O} &=&{}q_{s}^{O}\left( 1-q_{s}^{O}\right) \left[ \left( \left[
1-s_{s}^{I}(g,q)\right] -\left[ 1-s_{s}^{O}(g,q)\right] \frac{n\left(
1-\sum_{i}g_{i}q_{i}^{O}\right) }{K}\right) -\left( \Psi _{O}-\Psi
_{I}\right) \right] -\left( q_{s}^{O}\right) ^{2}\Phi \\
&&+\left[ \left( 1-q_{s}^{O}\right)
Z_{s}^{I,O}(g,q)-q_{s}^{O}Z_{s}^{O,I}(g,q)\frac{n\left(
1-\sum_{i}g_{i}q_{i}^{O}\right) }{K}\right] ,
\end{eqnarray*}%
Thus the differences between strategies in the state switching dynamics will
be described by survival and switching brackets. Note that after some
rearrangement the above equation can be presented in the form revealing
factors responsible for growth and decline:

\begin{eqnarray}
\dot{q}_{s}^{O} &=&{}\left( 1-q_{s}^{O}\right) \left[ q_{s}^{O}\left[
1-s_{s}^{I}(g,q)\right] +Z_{s}^{I,O}(g,q)\right] -q_{s}^{O}\left[ {}\left(
1-q_{s}^{O}\right) \left[ 1-s_{s}^{O}(g,q)\right] +Z_{s}^{O,I}(g,q)\right] 
\frac{n\left( 1-\sum_{i}g_{i}q_{i}^{O}\right) }{K}  \notag \\
&&-{}q_{s}^{O}\left( 1-q_{s}^{O}\right) \left( \Psi _{O}-\Psi _{I}\right)
-\left( q_{s}^{O}\right) ^{2}\Phi {},  \label{swO}
\end{eqnarray}%
\bigskip showing that for growth is responsible action chosen as Intruder
and its impact is described by the term $\mathbf{Inc}=\left( q_{s}^{O}\left[
1-s_{s}^{I}(g,q)\right] +Z_{s}^{I,O}(g,q)\right) $, while decline is
determined by action chosen as Owner and its impact is described by term $%
\mathbf{Dec}=\left( \left( 1-q_{s}^{O}\right) \left[ 1-s_{s}^{O}(g,q)\right]
+Z_{s}^{O,I}(g,q)\right) $. Then the switching dynamics (\ref{swO}) can be
denoted in the simplified form:\bigskip

\begin{equation}
\dot{q}_{s}^{O}={}\left( 1-q_{s}^{O}\right) \mathbf{Inc}-q_{s}^{O}\mathbf{Dec%
}\frac{n\left( 1-\sum_{i}g_{i}q_{i}^{O}\right) }{K}-{}q_{s}^{O}\left(
1-q_{s}^{O}\right) \left( \Psi _{O}-\Psi _{I}\right) -\left(
q_{s}^{O}\right) ^{2}\Phi .  \label{simO}
\end{equation}%
\bigskip Now we can derive the increase and decrease factors for Hawk and
Dove actions:%
\begin{eqnarray*}
\mathbf{Inc}_{H} &=&q_{s}^{O}H^{O}\left( 1-s\right) \frac{%
n\sum_{i}g_{i}q_{i}^{O}}{K}+1-H^{O}\left[ 1-0.5s\right] \frac{%
n\sum_{i}g_{i}q_{i}^{O}}{K} \\
&=&1+\left[ q_{s}^{O}\left( 1-s\right) -1+0.5s\right] H^{O}\frac{%
n\sum_{i}g_{i}q_{i}^{O}}{K} \\
\mathbf{Inc}_{D} &=&1-(1+H^{O})0.5\frac{n\sum_{i}g_{i}q_{i}^{O}}{K} \\
\mathbf{Dec}_{H} &=&\left[ \left( 1-q_{s}^{O}\right) \left( 1-s\right) +0.5s%
\right] H^{I} \\
\mathbf{Dec}_{D} &=&0.5(1+H^{I}).
\end{eqnarray*}
Note that $\mathbf{Inc}_{D}>\mathbf{Inc}_{H}$ when 
\begin{eqnarray*}
(1+H^{O})0.5 &<&\left[ 1-q_{s}^{O}\left( 1-s\right) -0.5s\right] H^{O} \\
0.5 &<&\left( 0.5-q_{s}^{O}\right) (1-s)H^{O}.
\end{eqnarray*}%
This condition is never satisfied. Similarly we have that $\mathbf{Dec}_{D}>%
\mathbf{Dec}_{H}$ when%
\begin{eqnarray*}
(1+H^{I})0.5 &>&\left[ \left( 1-q_{s}^{O}\right) \left( 1-s\right) +0.5s%
\right] H^{I} \\
0.5 &>&\left( 0.5-q_{s}^{O}\right) \left( 1-s\right) H^{I}.
\end{eqnarray*}

The second condition is always satisfied except the marginal case of $%
q_{s}^{O}=s=0$ and $H^{I}=1$. In effect Hawk action is most efficient for
switching in both roles. By substituting increase and decrease factors to (%
\ref{simO}) we can derive switching dynamics for respective strategies:%
\newline
Hawk strategy ($\mathbf{Inc}_{H}$ and $\mathbf{Dec}_{H}$)%
\begin{eqnarray*}
\dot{q}_{H}^{O} &=&{}\left( 1-q_{H}^{O}\right) \left( 1+\left[
q_{H}^{O}\left( 1-s\right) -1+0.5s\right] H^{O}\frac{n\sum_{i}g_{i}q_{i}^{O}%
}{K}\right) \\
&&-q_{H}^{O}\left[ \left( 1-q_{H}^{O}\right) \left( 1-s\right) +0.5s\right]
H^{I}\frac{n\left( 1-\sum_{i}g_{i}q_{i}^{O}\right) }{K} \\
&&-{}q_{H}^{O}\left( 1-q_{H}^{O}\right) \left( \Psi _{O}-\Psi _{I}\right)
-\left( q_{H}^{O}\right) ^{2}\Phi
\end{eqnarray*}%
Dove strategy ($\mathbf{Inc}_{D}$ and $\mathbf{Dec}_{D}$)%
\begin{eqnarray*}
\dot{q}_{D}^{O} &=&{}\left( 1-q_{D}^{O}\right) \left( 1-(1+H^{O})0.5\frac{%
n\sum_{i}g_{i}q_{i}^{O}}{K}\right) -q_{D}^{O}0.5(1+H^{I})\frac{n\left(
1-\sum_{i}g_{i}q_{i}^{O}\right) }{K} \\
&&-{}q_{D}^{O}\left( 1-q_{D}^{O}\right) \left( \Psi _{O}-\Psi _{I}\right)
-\left( q_{D}^{O}\right) ^{2}\Phi .
\end{eqnarray*}%
Bourgeois strategy ($\mathbf{Inc}_{D}$ and $\mathbf{Dec}_{H}$)%
\begin{eqnarray*}
\dot{q}_{B}^{O} &=&{}\left( 1-q_{B}^{O}\right) \left( 1-(1+H^{O})0.5\frac{%
n\sum_{i}g_{i}q_{i}^{O}}{K}\right) -q_{B}^{O}\left[ \left(
1-q_{B}^{O}\right) \left( 1-s\right) +0.5s\right] H^{I}\frac{n\left(
1-\sum_{i}g_{i}q_{i}^{O}\right) }{K} \\
&&-{}q_{B}^{O}\left( 1-q_{B}^{O}\right) \left( \Psi _{O}-\Psi _{I}\right)
-\left( q_{B}^{O}\right) ^{2}\Phi .
\end{eqnarray*}%
Antibourgeois strategy ($\mathbf{Inc}_{H}$ and $\mathbf{Dec}_{D}$)%
\begin{eqnarray*}
\dot{q}_{A}^{O} &=&{}\left( 1-q_{A}^{O}\right) \left( 1+\left[
q_{A}^{O}\left( 1-s\right) -1+0.5s\right] H^{O}\frac{n\sum_{i}g_{i}q_{i}^{O}%
}{K}\right) -q_{A}^{O}0.5(1+H^{I})\frac{n\left(
1-\sum_{i}g_{i}q_{i}^{O}\right) }{K} \\
&&-{}q_{A}^{O}\left( 1-q_{A}^{O}\right) \left( \Psi _{O}-\Psi _{I}\right)
-\left( q_{A}^{O}\right) ^{2}\Phi .
\end{eqnarray*}

\subsection{Mortality stage}

Average mortality rates are:

\begin{equation*}
\bar{D}_{H}(g,q,n)=\left[ q_{H}^{O}H^{I}\left(
1-\sum_{i}g_{i}q_{i}^{O}\right) +\left( 1-q_{H}^{O}\right)
H^{O}\sum_{i}g_{i}q_{i}^{O}\right] \frac{n}{K}(1-s)
\end{equation*}

\begin{equation*}
\bar{D}_{D}(g,q,n)=0
\end{equation*}%
\begin{equation*}
\bar{D}_{B}(g,q,n)=q_{B}^{O}H^{I}\left( 1-s\right) \frac{n\left(
1-\sum_{i}g_{i}q_{i}^{O}\right) }{K}
\end{equation*}%
\begin{equation*}
\bar{D}_{A}(g,q,n)=\left( 1-q_{A}^{O}\right) H^{O}\left( 1-s\right) \frac{%
n\sum_{i}g_{i}q_{i}^{O}}{K}.
\end{equation*}%
where $\bar{D}_{D}$ is smallest and $\bar{D}_{H}$ greatest, and $\bar{D}_{B}<%
\bar{D}_{A}$ when%
\begin{equation*}
\frac{q_{B}^{O}}{\sum_{H}g_{i}q_{i}^{O}}<\frac{\left( 1-q_{A}^{O}\right) }{%
\sum_{H}g_{i}\left( 1-q_{i}^{O}\right) }
\end{equation*}%
Average mortality rate of the whole population is:%
\begin{gather}
\bar{D}(g,q,n)=g_{H}\bar{D}_{H}(g,q,n)+g_{D}\bar{D}_{D}(g,q,n)+g_{B}\bar{D}%
_{B}(g,q,n)+g_{A}\bar{D}_{A}(g,q,n)  \notag \\
=\left[ \left( g_{H}q_{H}^{O}+g_{B}q_{B}^{O}\right) H^{I}\left( 1-s\right)
\left( 1-\sum_{i}g_{i}q_{i}^{O}\right) +\left( g_{H}\left(
1-q_{H}^{O}\right) +g_{A}\left( 1-q_{A}^{O}\right) \right) H^{O}\left(
1-s\right) \sum_{i}g_{i}q_{i}^{O}\right] \frac{n}{K}.  \label{avermort}
\end{gather}

Detailed derivation is in Appendix 1

\subsection{Selection dynamics}

Now let us describe the selection dynamics 
\begin{equation}
\dot{g}_{s}=g_{s}\left[ \left( \bar{\Phi}_{s}(q_{s}^{O})-\bar{\Phi}%
(g,q)\right) -\left( \bar{D}_{s}(g,q,n)-\bar{D}(g,q,n)\right) -\left( \bar{%
\Psi}_{s}(q_{s})-\bar{\Psi}(g,q)\right) \right] ,  \label{selection}
\end{equation}

which can be reduced to 
\begin{equation}
\dot{g}_{s}=g_{s}\left[ \left[ \Phi -\Psi _{O}+\Psi _{I}\right] \left(
q_{s}^{O}-\sum_{i}gq_{i}^{O}\right) -\left( \bar{D}_{s}(g,q,n)-\bar{D}%
(g,q,n)\right) \right] .  \label{sel}
\end{equation}%
Detailed derivation is in Appendix 2.\newline
The equation for population size will be: 
\begin{eqnarray*}
\dot{n} &=&n\left[ \bar{\Phi}(g,q)-\bar{\Psi}(g,q)-\bar{D}(g,q,n)\right] \\
&=&n\left[ \left[ \Phi -\Psi _{O}+\Psi _{I}\right] \sum_{i}g_{i}q_{i}^{O}-%
\Psi _{I}-\bar{D}(g,q,n)\right] .
\end{eqnarray*}%
Impact of the population size $n$ on function $\bar{D}(g,q,n)$ is realized
by multiplicative term $\dfrac{n}{K}$. Therefore, the average mortality (\ref%
{avermort}) can be presented as $\bar{D}(g,q,n)=\tilde{D}(g,q)\dfrac{n}{K}$,
in effect attractor population size can be calculated:%
\begin{equation*}
\tilde{n}=K\frac{\left[ \Phi -\Psi _{O}+\Psi _{I}\right]
\sum_{i}g_{i}q_{i}^{O}-\Psi _{I}}{\tilde{D}(g,q)}
\end{equation*}%
Therefore (\ref{sel}) contains two factors, first term $\left(
q_{s}^{O}-\sum_{i}gq_{i}^{O}\right) $ weighted by background payoffs bracket
describes differences in role distributions of the subpopulations of
carriers, and the second term $\left( \bar{D}_{s}(g,q,n)-\bar{D}%
(g,q,n)\right) $ describes differences in average mortalities resulting from
role distributions too, but also from individual actions determined by
strategies. Thus it mixes the results of strategies used in the focal
interaction and background payoffs. In addition level of individual
interactions is determined by actual compositions of the carrier
subpopulations and we have complex trade-offs between those levels. This is
even more complicated situation than in the case of the sex ratio evolution
(Argasinski 2012, 2013, 2018) where selection is determined by sex ratios in
carrier subpopulations (thus subpopulation compositions not individual
traits) while individual actions are responsible for the adjustment of those
sex ratios leading to the double level selection system. However, the
difference between those systems is that in the case of sex ratio evolution
adjustment of the subpopulation composition is realized by demographic
process (differences in the numbers of births) while in Owner-Intruder game
by switching dynamics independent from demography. Therefore, it cannot be
regarded as the multi-level selection, even if the subpopulation composition
plays important role in this process. This is rather related to the extended
phenotype concept (Dawkins 2016), since the subpopulation composition is the
direct result of the behaviour (i.e. leaving the property or not).\newpage

\subsection{Obtained system}

Summarizing the above derivations we have the system 
\begin{eqnarray}
\dot{q}_{H}^{O} &=&{}\left( 1-q_{H}^{O}\right) \left( 1+\left[
q_{H}^{O}\left( 1-s\right) -1+0.5s\right] H^{O}\frac{n\sum_{i}g_{i}q_{i}^{O}%
}{K}\right)  \notag \\
&&-q_{H}^{O}\left[ \left( 1-q_{H}^{O}\right) \left( 1-s\right) +0.5s\right]
H^{I}\frac{n\left( 1-\sum_{i}g_{i}q_{i}^{O}\right) }{K}  \notag \\
&&-{}q_{H}^{O}\left( 1-q_{H}^{O}\right) \left( \Psi _{O}-\Psi _{I}\right)
-\left( q_{H}^{O}\right) ^{2}\Phi
\end{eqnarray}

\begin{eqnarray}
\dot{q}_{D}^{O} &=&{}\left( 1-q_{D}^{O}\right) \left( 1-(1+H^{O})0.5\frac{%
n\sum_{i}g_{i}q_{i}^{O}}{K}\right) -q_{D}^{O}0.5(1+H^{I})\frac{n\left(
1-\sum_{i}g_{i}q_{i}^{O}\right) }{K}  \notag \\
&&-{}q_{D}^{O}\left( 1-q_{D}^{O}\right) \left( \Psi _{O}-\Psi _{I}\right)
-\left( q_{D}^{O}\right) ^{2}\Phi .
\end{eqnarray}

\begin{eqnarray}
\dot{q}_{B}^{O} &=&{}\left( 1-q_{B}^{O}\right) \left( 1-(1+H^{O})0.5\frac{%
n\sum_{i}g_{i}q_{i}^{O}}{K}\right) -q_{B}^{O}\left[ \left(
1-q_{B}^{O}\right) \left( 1-s\right) +0.5s\right] H^{I}\frac{n\left(
1-\sum_{i}g_{i}q_{i}^{O}\right) }{K}  \notag \\
&&-{}q_{B}^{O}\left( 1-q_{B}^{O}\right) \left( \Psi _{O}-\Psi _{I}\right)
-\left( q_{B}^{O}\right) ^{2}\Phi .
\end{eqnarray}

\begin{eqnarray}
\dot{q}_{A}^{O} &=&{}\left( 1-q_{A}^{O}\right) \left( 1+\left[
q_{A}^{O}\left( 1-s\right) -1+0.5s\right] H^{O}\frac{n\sum_{i}g_{i}q_{i}^{O}%
}{K}\right) -q_{A}^{O}0.5(1+H^{I})\frac{n\left(
1-\sum_{i}g_{i}q_{i}^{O}\right) }{K}  \notag \\
&&-{}q_{A}^{O}\left( 1-q_{A}^{O}\right) \left( \Psi _{O}-\Psi _{I}\right)
-\left( q_{A}^{O}\right) ^{2}\Phi .
\end{eqnarray}%
\bigskip

\begin{equation}
\dot{g}_{D}=g_{D}\left[ \left[ \Phi -\Psi _{O}+\Psi _{I}\right] \left(
q_{D}^{O}-\sum_{i}g_{i}q_{i}^{O}\right) +\tilde{D}(g,q)\dfrac{n}{K}\right]
\end{equation}%
\begin{equation}
\dot{g}_{B}=g_{B}\left[ \left[ \Phi -\Psi _{O}+\Psi _{I}\right] \left(
q_{B}^{O}-\sum_{i}g_{i}q_{i}^{O}\right) -\left( q_{B}^{O}H^{I}\left(
1-s\right) \frac{n\left( 1-\sum_{i}g_{i}q_{i}^{O}\right) }{K}-\tilde{D}(g,q)%
\dfrac{n}{K}\right) \right]
\end{equation}%
\begin{equation}
\dot{g}_{A}=g_{A}\left[ \left[ \Phi -\Psi _{O}+\Psi _{I}\right] \left(
q_{A}^{O}-\sum_{i}g_{i}q_{i}^{O}\right) -\left( \left( 1-q_{A}^{O}\right)
H^{O}\left( 1-s\right) \frac{n\sum_{i}g_{i}q_{i}^{O}}{K}-\tilde{D}(g,q)%
\dfrac{n}{K}\right) \right]
\end{equation}%
\begin{eqnarray}
\dot{n} &=&n\left[ \bar{\Phi}(g,q)-\bar{\Psi}(g,q)-\bar{D}(g,q)\right] 
\notag \\
&=&n\left[ \left[ \Phi -\Psi _{O}+\Psi _{I}\right] \sum_{i}g_{i}q_{i}^{O}-%
\Psi _{I}-\tilde{D}(g,q)\dfrac{n}{K}\right]
\end{eqnarray}%
where%
\begin{eqnarray*}
\tilde{D}(g,q) &=&\left[ \left( g_{H}q_{H}^{O}+g_{B}q_{B}^{O}\right)
H^{I}\left( 1-s\right) \left( 1-\sum_{i}g_{i}q_{i}^{O}\right) +\left(
g_{H}\left( 1-q_{H}^{O}\right) +g_{A}\left( 1-q_{A}^{O}\right) \right)
H^{O}\left( 1-s\right) \sum_{i}g_{i}q_{i}^{O}\right] \\
H^{O} &=&\dfrac{\sum_{H}g_{i}q_{i}^{O}}{\sum_{i}g_{i}q_{i}^{O}}\text{ \ \
and \ \ \ }H^{I}=\dfrac{\sum_{H}g_{i}\left( 1-q_{i}^{O}\right) }{%
\sum_{i}g_{i}\left( 1-q_{i}^{O}\right) },
\end{eqnarray*}%
\bigskip and the attracting density surface is: \bigskip

\begin{equation*}
\tilde{n}=K\frac{\left[ \Phi -\Psi _{O}+\Psi _{I}\right]
\sum_{i}g_{i}q_{i}^{O}-\Psi _{I}}{\tilde{D}(g,q)}
\end{equation*}

\subsection{Resulting static fitness measure}

From selection dynamics (\ref{sel}) we can derive proper static fitness
function which is 
\begin{equation*}
F_{s}(g,q)=q_{s}^{O}\left[ \Phi -\Psi _{O}+\Psi _{I}\right] -\bar{D}%
_{s}(g,q,n).
\end{equation*}%
Therefore selection is driven by two factors, focal interaction average
mortality $\bar{D}_{s}(g,q,n)$ and the differences in background vital rates
caused by different role allocations. The second stage is described by
factor $q_{s}^{O}\left[ \Phi -\Psi _{O}+\Psi _{I}\right] $. Note that role
allocation affects the value of $\bar{D}_{s}(g,q,n)$ too. In the expanded
form above function can be presented as%
\begin{equation*}
F_{s}(g,q)=q_{s}^{O}\left( \Phi -\Psi _{O}+\Psi _{I}\right) -\left(
q_{s}^{O} \left[ 1-s_{s}^{O}(g,q)\right] \frac{n\left(
1-\sum_{i}g_{i}q_{i}^{O}\right) }{K}+\left( 1-q_{s}^{O}\right) \left[
1-s_{s}^{I}(g,q)\right] \right)
\end{equation*}

substitution of the respective action specific survival payoffs (\ref{survOH}%
), (\ref{survIH}) and (\ref{survD}) for $s_{s}^{O}(g,q)$ and $s_{s}^{I}(g,q)$
leads to

\begin{eqnarray*}
F_{H}(g,q) &=&q_{H}^{O}\left( \Phi -\Psi _{O}+\Psi _{I}\right) -\left[
q_{H}^{O}H^{I}\left( 1-\sum_{i}g_{i}q_{i}^{O}\right) +\left(
1-q_{H}^{O}\right) H^{O}\sum_{i}g_{i}q_{i}^{O}\right] \frac{n}{K}\left(
1-s\right) \\
F_{D}(g,q) &=&q_{D}^{O}\left( \Phi -\Psi _{O}+\Psi _{I}\right) \\
F_{B}(g,q) &=&q_{B}^{O}\left( \Phi -\Psi _{O}+\Psi _{I}\right)
-q_{B}^{O}\left( 1-\sum_{i}g_{i}q_{i}^{O}\right) H^{I}\frac{n}{K}\left(
1-s\right) \\
F_{A}(g,q) &=&q_{A}^{O}\left( \Phi -\Psi _{O}+\Psi _{I}\right) -\left(
1-q_{s}^{O}\right) \sum_{i}g_{i}q_{i}^{O}H^{O}\frac{n}{K}\left( 1-s\right)
\end{eqnarray*}%
\bigskip \bigskip

\section{Numerical simulations}

Due to complexity of the obtained model, to avoid the inflation of the
paper, we will limit ourselves to numerical simulations, leaving the
detailed analysis for the separate paper. First important thing is the
choice of the biologically relevant values of parameters. Let us focus on
the background mortality and fertility rates $\Psi _{O}$, $\Psi _{I}$ and $%
\Phi $. The timescale of the model is adjusted to the constant Intruder's
per capita intensity of nest site inspection (equivalent to the interaction
rate with possible Owner) which leads to (\ref{tauratio}). Therefore,
realistic situation is when Owners background mortality $\Psi _{O}$ is
significantly smaller than respective parameter for Intruders $\Psi _{I}$.
Fertility rate $\Phi $ should be greater than average mortality to avoid the
extinction of the population. Inspections of the nest sites occur in the
behavioral timescale and are more frequent than births and background
deaths. This will allow Intruders to do few inspections before death.
Therefore, background vital rates should be smaller than focal game
interaction rate, leading to $\Psi _{O}<\Psi _{I}<\Phi <1$. In those cases
new model reproduces the predictions that are known from literature.
Competition between pure Hawk and pure Dove leads to the mixed equilibrium
which depends on the survival of the fight between Hawks. This situation is
shown on Fig.4, for very low fight survival probability $s=0,001$ and the
background mortality and fertility rates $\Psi _{O}=0.01$, $\Psi _{I}=0.2$
and $\Phi =0.05$. What is interesting, for such harsh conditions the stable
frequency of hawks is relatively high, in addition the proportion of Owners
among Hawks is close to 1, while for Doves is slightly lower. In effect
subpopulation of homeless Intruders contains majority of Doves. Another
interesting observation is that the total population size is smaller than
the number of nest sites $K$. Therefore, in the environment is enough nest
sites for all individuals. This means that the availability of nest sites is
not the main limiting factor here. The suppression is driven by pressure of
the background mortality related to the searching for free nest site.
Simulations also show that both pure strategies are outcompeted by Bourgeois
and paradoxical Antibourgeois strategies. Both those strategies can
successfully invade the population and be evolutionarily stable. In the
population composed of competing Bourgeois and Antibourgeois individuals,
both strategies are stable and invasion barrier is exactly 0.5. However,
when we add some fraction of Hawks or Doves, the invasion barrier will shift
closer to the Antibourgeois pure state, leading to the increase of the basin
of attraction of Bourgeois. For some parameter values it is possible another
interesting situation, when Bourgeois is not stable but exists mixed
equilibrium composed of Bourgeois and small fraction of pure Hawks. Let us
focus on this situation. Assume that number of nest sites is $K=10000$. We
can choose values $\Psi _{O}=0.01$, $\Psi _{I}=0.2$ and $\Phi =0.8$ and the
survival of the fight as $s=0.4$. For initial state of the population we
assume that $n(0)=70$ and all for all strategies role distributions are
equal to $0.1$ and gene frequencies to $0.25$. The trajectories of
population parameters are depicted on Fig.5. In addition, Fig.6 shows
changes in the average mortalities $\bar{D}_{s}(g,q,n)$ for all strategies
and frequencies of Hawks among Owners and Intruders $H^{O}$ and $H^{I}$,
which determine those mortalities. They show the increase in mortality
associated with population growth. This is caused by increasing number of
fights related to the increase of the fraction of the occupied nest sites.%

\begin{figure}[h!]
\centering
\includegraphics[width=14cm]{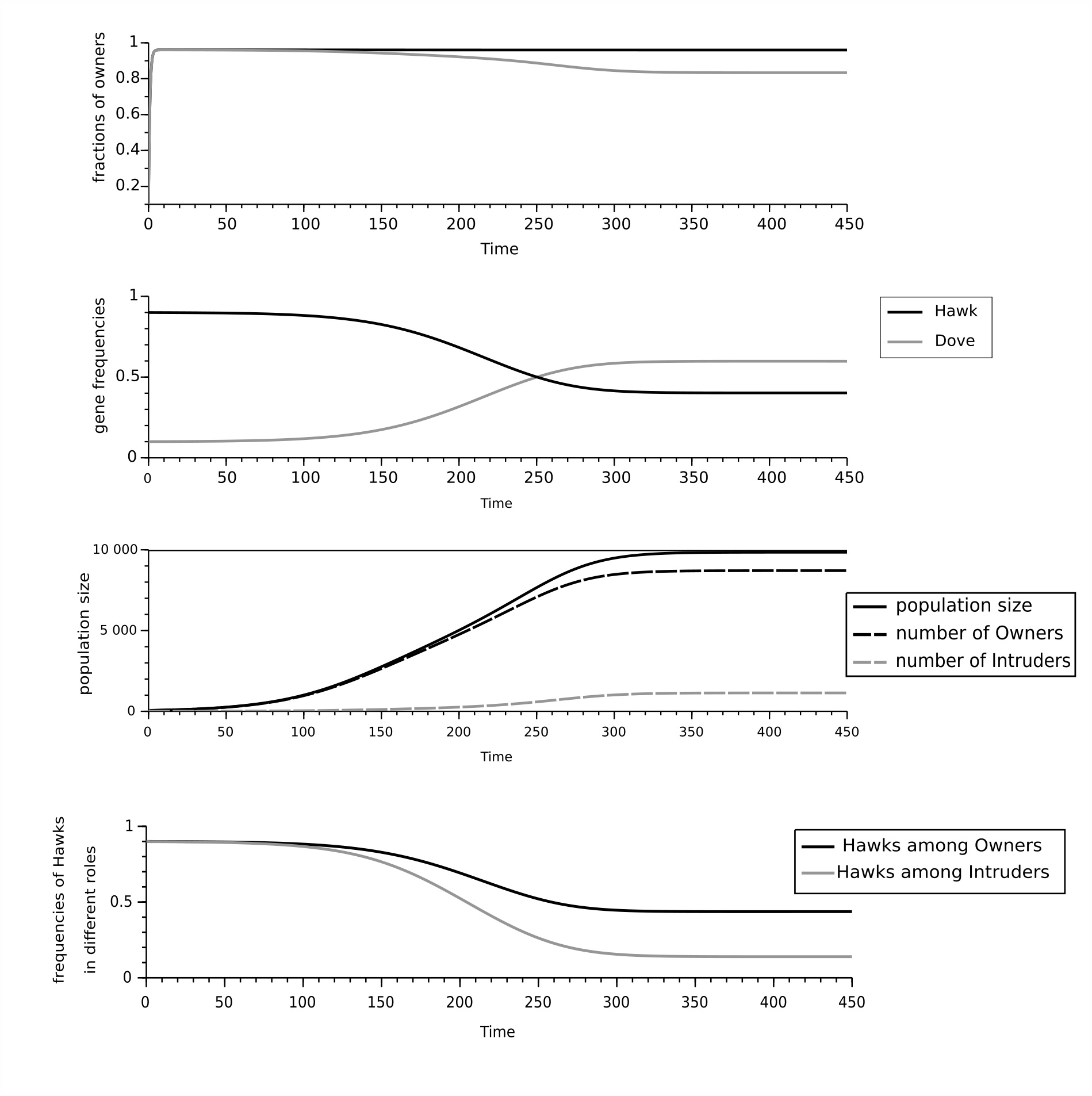}
\caption{Case of competing pure Hawk and
Dove strategies. Plots of role distributions $q_{s}^{O}$, strategy
frequencies $g_{s}$ and numbers of Owners $n\sum_{s}g_{s}q_{s}^{O}$,
Intruders $n\left( 1-\sum_{s}g_{s}q_{s}^{O}\right) $ and the population size 
$n$. Last panel shows proportions of Hawks among Owners and Intruders.
Proportion of Owners among Hawks is close to 1 while among doves is sligtly
lower. On the other hand among Intruders we have majority of Doves. In
effect among reproducing Owners thie proportion of Hawks is slightly greater
than Hawk gene frequency. Interesting is that the stable population sizeis
lower than the number of nest sites, thus this is not a lomiting factor in
this case. Suppression is rather driven by background mortalities.}
\end{figure}

\begin{figure}[h!]
\centering
\includegraphics[width=14cm]{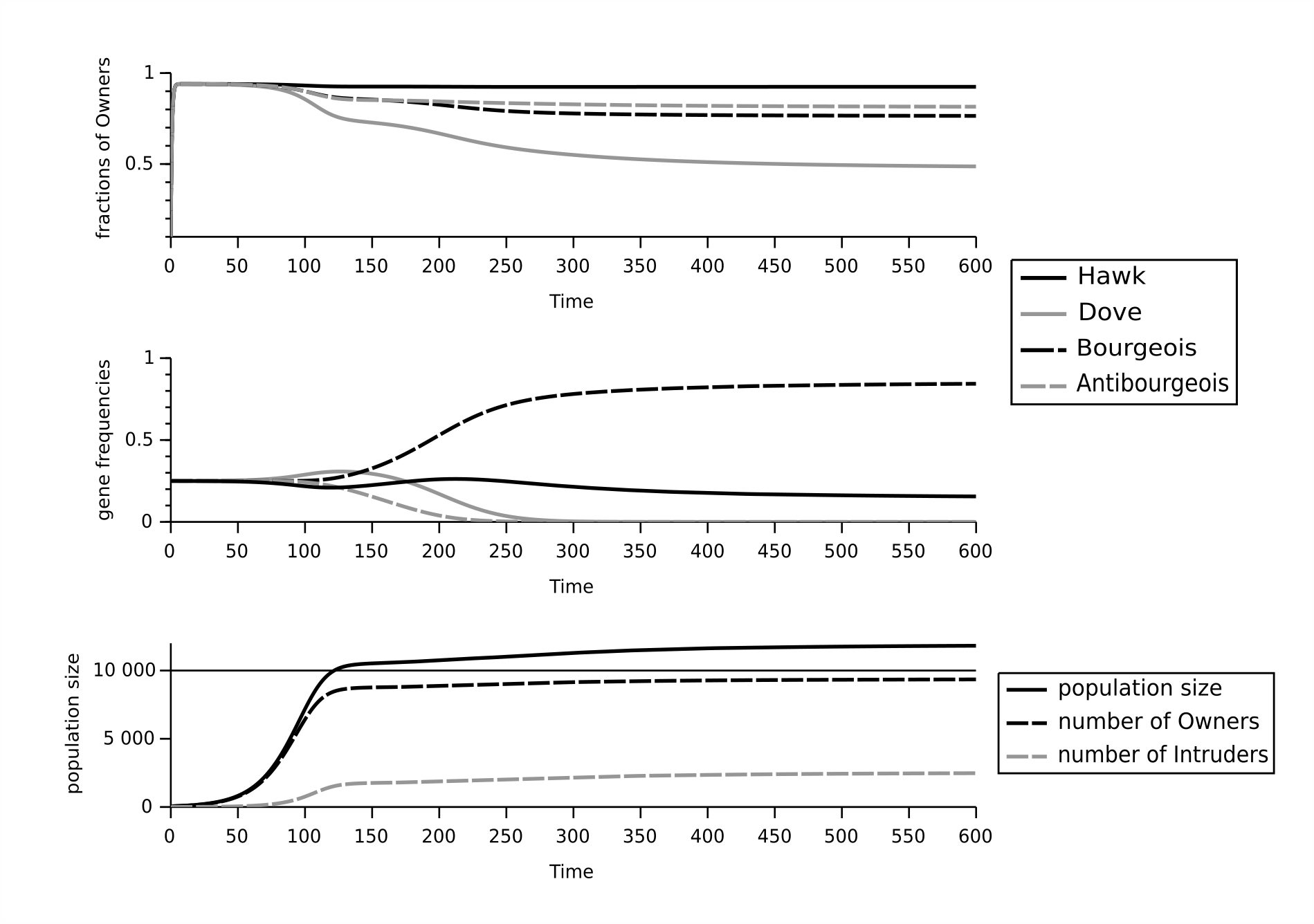}
\caption{Plots of role distributions $q_{s}^{O}$, strategy frequencies $g_{s}$ and numbers of
Owners $n\sum_{s}g_{s}q_{s}^{O}$, Intruders $n\left(1-\sum_{s}g_{s}q_{s}^{O}\right) $ and the population size $n$. Interesting is that the system converged to the mixture of Bourgeois and pure Hawks}
\end{figure}

\begin{figure}[h!]
\centering
\includegraphics[width=14cm]{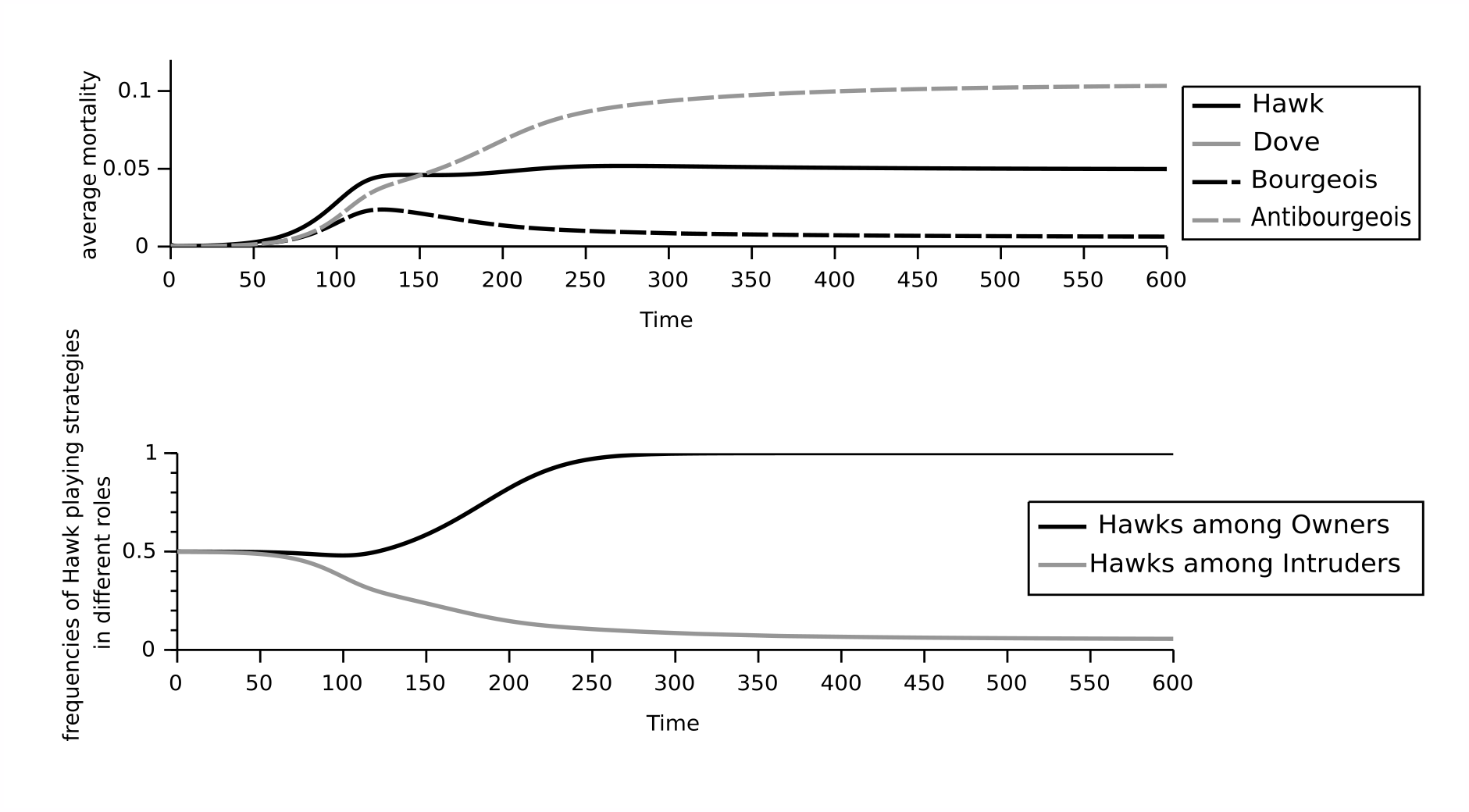}
\caption{Plots of changes in average mortality rates $\bar{D}_{s}$ for all
strategies and frequencies of Hawk playing strategies among Owners and
Intruders $H^{O}$ and $H^{I}$. Note that mortality increases with the
population growth since fights become more frequent.}
\end{figure}

\section{Discussion\protect\bigskip}

In this paper we introduced the combination of the dynamic evolutionary
games with the state based approach. We presented the system of differential
equations which describes the dynamics of state changes and completes the
replicator dynamics. Our new framework can be useful in the integration of
many distinct fields of theoretical evolutionary biology into coherent
synthetic theory. We illustrated this by examples of application of the new
framework to the classical problems.

The natural processes such as aging or the developmental process can be
described in terms of state changing processes. It can be useful for
formalization of the game theoretic notions such as "costs" and "benefits"
in terms of the stages of a life cycle of the individual. We presented a
simple attempt to this problem which shows the complexity of the phenomenon.
This example shows that the new framework presented in this paper can be
used in life history modelling (Stearns 1992, Roff 2002)\ The future
research on this topic should investigate the relationships of this type of
models with results related to the classical demography.

Example of dynamic owner-intruder game shows the complexity of the process
underlying the changes of life situation determined by strategy.\ It clearly
shows that the assumption in the classical models (Maynard Smith 1982) that
the individual can be owner or intruder with the probability 0.5 was very
strong simplification. Owner-Intruder example emphasizes the importance of
the definition of the demographic vital rates as the products of the
interaction rates and the demographic outcomes of the interaction
(interpreted as the game round, Argasinski and Broom 2018a). In this example
different interaction rates for opposite roles should be explicitly
considered in the model (for example in the average mortalities) thus this
problem cannot be clearly described in terms of abstract fitnesses expressed
in units of instantaneous growth rates. The case of stable mixed
Bourgeois/Hawk state is important from the point of view of the latest
theories on the impact of density dependence on the selection mechanisms,
which are one of important questions (Da\'{n}ko et al 2018). Latest Nest
Site Lottery models (Argasinski and Broom 2013b, Argasinski and Rudnicki
2017, Rudnicki 2018, Argasinski and Rudnicki submitted) silently\ assume
that all newborns competing for free nest sites are Bourgeois. Our new model
shows that this may be not the case. Some newborns can be Hawks and can try
to attack the occupied nest sites. Presence of small but significant
fraction of Hawks may alter the selection mechanism. Note that our model
describes very specific conflict for supply which are nest sites. Conflict
for other supplies such as food will have different structure and will be
driven by different mechanisms, for example models of kleptoparasitism
describe situations when ownership is not respected (Broom and Ruxton 1998;
Broom and Rychtar 2007,2009,2011; Luther et al 2007; Broom et al 2009) This
suggests that there is no general explanation for ownership respect and
different forms of ownership are shaped by different selection mechanisms.
Therefore we can expect different behavioral patterns for those distinct
conflicts, which means that the respect for nest site may not imply respect
for ownership of food and vice versa.

Important reason shown by new approach is the importance of the state
distribution for the determination of fitness, which is the quantity related
to the population growth rate (Metz 2008) but can be formalized in many ways
(Roff 2009). The dynamic equilibria of the state changing process,
constituting the role/state distributions, should be explicitly taken into
consideration in the fitness evaluation. The possible perturbations of the
stable distribution of states should be incorporated to the modern
development of the ESS theory to increase its consistence (Houston and
McNamara 2005, McNamara 2013) .\ This will be interesting contribution to
theoretical studies related to the individual level, originated by \L %
omnicki (1988), and in particular related to research on animal
personalities (Dall et al. 2004; Wolf et al. 2007; Wolf and Weissing 2010,
Wolf and Weissing 2012; Wolf and McNamara 2012).\bigskip

We want to thank John McNamara for scientific mentoring, helpful suggestions
and support during scientific visits. In addition, we want to thank Mark
Broom and Jan Koz\l owski a for their support of the project.\ This paper
was supported by the Polish National Science Centre Grant No.
2013/08/S/NZ8/00821 FUGA2 (KA) and Grant No. 2017/27/B/ST1/00100 OPUS (RR)

\bigskip

\subsection*{References\newline
}

Argasinski K. (2006) Dynamic multipopulation and density dependent
evolutionary games related to replicator dynamics. A metasimplex concept.
Math Biosci 202 88-114. \newline
Argasinski K. (2012) The dynamics of sex ratio evolution Dynamics of global
population parameters. J Theor Biol 309 134-146.\newline
Argasinski K. (2013) The Dynamics of Sex Ratio Evolution: From the Gene
Perspective to Multilevel Selection. PloS ONE, 8(4), e60405\newline
Argasinski K. (2018) The dynamics of sex ratio evolution: the impact of
males as passive gene carriers on multilevel selection. Dynamic Games and
Applications, 8(4), 671-695.\newline
Argasinski K. Broom M. (2013a)\ Ecological theatre and the evolutionary
game: how environmental and demographic factors determine payoffs in
evolutionary games. J Math Biol. 1;67(4):935-62. \newline
Argasinski K. Broom M.(2013b) The nest site lottery: how selectively neutral
density dependent growth suppression induces frequency dependent selection.
Theor Pop Biol 90 82-90. \newline
Argasinski, K. Broom M. (2018a). Interaction rates, vital rates, background
fitness and replicator dynamics: how to embed evolutionary game structure
into realistic population dynamics. Theory in Biosciences, 137(1), 33-50.%
\newline
Argasinski K. Broom M. (2018b). Evolutionary stability under limited
population growth: Eco-evolutionary feedbacks and replicator dynamics.
Ecological Complexity, 34, 198-212.\newline
Argasinski K, Rudnicki R. (2017). Nest site lottery revisited: Towards a
mechanistic model of population growth suppressed by the availability of
nest sites. J Theor Biol, 420, 279-289.\newline
Cressman R. (1992) The Stability Concept of Evolutionary Game Theory.
Springer. \newline
Cressman R., \& K\v{r}ivan, V. (2019). Bimatrix games that include
interaction times alter the evolutionary outcome: The Owner--Intruder game.
Journal of theoretical biology, 460, 262-273.\newline
Da\'{n}ko MJ, Burger O, Argasi\'{n}ski K, \& Koz\l owski, J (2018).
Extrinsic mortality can shape life-history traits, including senescence.
Evolutionary biology, 45(4), 395-404.\newline
Broom M., Ruxton G. D. (1998). Evolutionarily stable stealing: game theory
applied to kleptoparasitism. Annals of Human Genetics, 62(5), 453-464.%
\newline
Broom, M., Rycht\'{a}\v{r}, J. (2007). The evolution of a kleptoparasitic
system under adaptive dynamics. Journal of Mathematical Biology, 54(2),
151-177.\newline
Broom, M., Rycht\'{a}\v{r}, J. (2009). A game theoretical model of
kleptoparasitism with incomplete information. Journal of mathematical
biology, 59(5), 631-649.\newline
Broom, M., Luther, R. M., Rycht\'{a}r, J. (2009). A hawk-dove game in
kleptoparasitic populations. Journal of Combinatorics, Information and
System Sciences, 4, 449-462.\newline
Broom, M., Rycht\'{a}\v{r}, J. (2011). Kleptoparasitic melees---modelling
food stealing featuring contests with multiple individuals. Bulletin of
mathematical biology, 73(3), 683-699.\newline
Brunetti, I., Hayel, Y., \& Altman, E. (2015). State policy couple dynamics
in evolutionary games. In 2015 American Control Conference (ACC) (pp.
1758-1763). IEEE.\newline
Brunetti, I., Hayel, Y., \& Altman, E. (2018). State-policy dynamics in
evolutionary games. Dynamic Games and Applications, 8(1), 93-116. \newline
Dawkins R (2016).The extended phenotype: The long reach of the gene. Oxford
University Press.\newline
Eshel I, Sansone E. (1995). Owner-intruder conflict, Grafen effect and
self-assessment. The bourgeois principle re-examined. J. Theor Biol, 177(4),
341-356.\newline
Grafen A. (2006) Optimization of inclusive fitness. Journal of Theoretical
Biology. Feb 7;238(3):541-63.\newline
Grafen A. (1987). The logic of divisively asymmetric contests: respect for
ownership and the desperado effect. Anim. Behav. 35, 462--467.\newline
Hofbauer J, Sigmund.K (1988) The Theory of Evolution and Dynamical Systems.
Cambridge University Press. \newline
Hofbauer J, Sigmund.K (1998) Evolutionary Games and Population Dynamics.
Cambridge University Press. \newline
Houston AI, McNamara J (1999) JM. Models of adaptive behaviour: an approach
based on state. Cambridge University Press; \newline
Houston AI, McNamara J (2005). John Maynard Smith and the importance of
consistency in evolutionary game theory. Biol Phil 20(5) 933-950. \newline
Huang W, Hauert C, Traulsen A (2015). Stochastic game dynamics under
demographic fluctuations. PNAS, 112(29), 9064-9069\newline
Hui, C. (2006). Carrying capacity, population equilibrium, and environment's
maximal load. Ecological Modelling, 192(1-2), 317-320.\newline
Hauert C, Holmes M. Doebeli M (2006) Evolutionary games and population
dynamics: maintenance of cooperation in public goods games. Proceedings of
the Royal Society B: Biological Sciences, 273(1600), pp.2565--2570.\newline
Hauert C, Wakano JY, Doebeli M, (2008) Ecological public goods games:
cooperation and bifurcation. Theor Pop Biol, 73(2), pp.257--263\newline
Leimar, O., \& Enquist, M. (1984). Effects of asymmetries in owner-intruder
conflicts. Journal of theoretical Biology, 111(3), 475-491.\newline
Kokko H, L\'{o}pez-Sepulcre, A Morrell LJ (2006). From hawks and doves to
self-consistent games of territorial behavior. The American Naturalist,
167(6), 901-912\newline
Kokko H, (2013) Dyadic contests: Modelling figths between two individuals,
in: Animal contests Hardy, I. C., \& Briffa, M. (Eds.) Cambridge University
Press\newline
Korona, R. (1989). Evolutionarily stable strategies in competition for
resource intake rate maximization. Behavioral ecology and sociobiology,
25(3), 193-199\newline
Korona, R. (1991). On the role of age and body size in risky animal
contests. Journal of theoretical biology, 152(2), 165-176\newline
K\v{r}ivan V, Galanthay TE, \& Cressman R. (2018). Beyond replicator
dynamics: From frequency to density dependent models of evolutionary games.
J. Theor. Biol. , 455, 232-248.\newline
Luther, R. M., Broom, M., \& Ruxton, G. D. (2007). Is food worth fighting
for? ESS's in mixed populations of kleptoparasites and foragers. Bulletin of
mathematical biology, 69(4), 1121-1146.\newline
\L omnicki A (1988). Population ecology of individuals. Princeton University
Press, Princeton, New Jersey. \newline
Maynard Smith J (1982) Evolution and the Theory of Games. Cambridge
University Press, Cambbridge, United Kingdom. \newline
McElreath, R., \& Boyd, R. (2008). Mathematical models of social evolution:
A guide for the perplexed. University of Chicago Press.\newline
McNamara JM (2013). Towards a richer evolutionary game theory. J Roy Soc
Interface 10(88) 20130544. \newline
Mylius, S. D. (1999). What pair formation can do to the battle of the sexes:
towards more realistic game dynamics. J. Theor. Biol, 197(4), 469-485.%
\newline
Roff DA (2008). Defining fitness in evolutionary models. J Genet 87,
339--348. \newline
Roff DA (2002) Life history evolution. Sinauer\newline
Rudnicki R (2017). Does a population with the highest turnover coefficient
win competition?. Journal of Difference Equations and Applications, 23(9),
1529-1541.\newline
Sherratt TN, Mesterton-Gibbons M. (2015). The evolution of respect for
property. Journal of evolutionary biology, 28(6), 1185-1202.\newline
Stearns SC.(1992) The evolution of life histories. Oxford: Oxford University
Press\newline
Metz JAJ (2008). Fitness. In: J\o rgensen, S.E., Fath, B.D. (Eds.),
Evolutionary Ecology. In: Encyclopedia of Ecology, vol. 2. Elsevier, pp.
1599--1612. \newline
Taylor PD. (1996) Inclusive fitness arguments in genetic models of
behaviour. Journal of mathematical biology. May 1;34(5-6):654-74.\newline
Van Veelen M. (2009) Group selection, kin selection, altruism and
cooperation: when inclusive fitness is right and when it can be wrong.
Journal of Theoretical Biology. Aug 7;259(3):589-600\newline
Wolf M., McNamara JM. (2012). On the evolution of personalities via
frequency-dependent selection. Am Nat 179(6), 679-692.\newline
Wolf M, Van Doorn GS, Leimar O, Weissing FJ. (2007). Life-history trade-offs
favour the evolution of animal personalities. Nature, 447(7144), 581-584.%
\newline
Wolf M, Weissing FJ. (2010). An explanatory framework for adaptive
personality differences. Phil Trans Proc Roy Soc B 365(1560), 3959-3968.%
\newline
Wolf M, Weissing FJ. (2012). Animal personalities: consequences for ecology
and evolution. Trends Ecol Evol 27(8), 452-461.\newline
Zhang F, Hui C (2011) Eco-Evolutionary Feedback and the Invasion of
Cooperation in Prisoner's Dilemma Games. PLoS ONE 6(11) e27523. \newline
doi:10.1371/journal.pone.0027523\bigskip

\section*{Appendix 1 Derivation of mortality rates}

Average mortality rates for respective strategies are%
\begin{eqnarray*}
\bar{D}_{H}(g,q,n) &=&q_{H}^{O}\left[ 1-s_{H}^{O}(g,q)\right] \frac{n\left(
1-\sum_{i}g_{i}q_{i}^{O}\right) }{K}+\left( 1-q_{H}^{O}\right) \left[
1-s_{H}^{I}(g,q)\right] = \\
&=&q_{H}^{O}H^{I}\left( 1-s\right) \frac{n\left(
1-\sum_{i}g_{i}q_{i}^{O}\right) }{K}+\left( 1-q_{H}^{O}\right) H^{O}\left(
1-s\right) \frac{n\sum_{i}g_{i}q_{i}^{O}}{K} \\
&=&\left[ q_{H}^{O}H^{I}\left( 1-\sum_{i}g_{i}q_{i}^{O}\right) +\left(
1-q_{H}^{O}\right) H^{O}\sum_{i}g_{i}q_{i}^{O}\right] \frac{n}{K}(1-s)
\end{eqnarray*}

\begin{equation*}
\bar{D}_{D}(g,q,n)=0
\end{equation*}%
\begin{equation*}
\bar{D}_{B}(g,q,n)=q_{B}^{O}\left[ 1-s_{H}^{O}(g,q)\right] \frac{n\left(
1-\sum_{i}g_{i}q_{i}^{O}\right) }{K}+\left( 1-q_{B}^{O}\right) \left[
1-s_{D}^{I}(g,q)\right] \\
=q_{B}^{O}H^{I}\left( 1-s\right) \frac{n\left(
1-\sum_{i}g_{i}q_{i}^{O}\right) }{K}.
\end{equation*}%
\begin{eqnarray*}
\bar{D}_{A}(g,q,n) &=&q_{A}^{O}\left[ 1-s_{D}^{O}(g,q)\right] \frac{n\left(
1-\sum_{i}g_{i}q_{i}^{O}\right) }{K}+\left( 1-q_{A}^{O}\right) \left[
1-s_{H}^{I}(g,q)\right] = \\
&=&\left( 1-q_{A}^{O}\right) H^{O}\left( 1-s\right) \frac{%
n\sum_{i}g_{i}q_{i}^{O}}{K}.
\end{eqnarray*}%
Obviously $\bar{D}_{D}$ is smallest and $\bar{D}_{H}$ greatest. We have that 
$\bar{D}_{B}<\bar{D}_{A}$ when%
\begin{eqnarray*}
q_{B}^{O}H^{I}\left( 1-s\right) \frac{n\left(
1-\sum_{i}g_{i}q_{i}^{O}\right) }{K} &<&\left( 1-q_{A}^{O}\right)
H^{O}\left( 1-s\right) \frac{n\sum_{i}g_{i}q_{i}^{O}}{K} \\
q_{B}^{O}\left( 1-\sum_{H}g_{i}q_{i}^{O}\right) &<&\left( 1-q_{A}^{O}\right)
\sum_{H}g_{i}q_{i}^{O} \\
\frac{q_{B}^{O}}{\sum_{H}g_{i}q_{i}^{O}} &<&\frac{\left( 1-q_{A}^{O}\right) 
}{\sum_{H}g_{i}\left( 1-q_{i}^{O}\right) }
\end{eqnarray*}%
Average mortality rate of the whole population is:%
\begin{gather*}
\bar{D}(g,q,n)=g_{H}\bar{D}_{H}(g,q,n)+g_{D}\bar{D}_{D}(g,q,n)+g_{B}\bar{D}%
_{B}(g,q,n)+g_{A}\bar{D}_{A}(g,q,n) \\
=g_{H}\left[ q_{H}^{O}H^{I}\left( 1-s\right) \frac{n\left(
1-\sum_{i}g_{i}q_{i}^{O}\right) }{K}+\left( 1-q_{H}^{O}\right) H^{O}\left(
1-s\right) \frac{n\sum_{i}g_{i}q_{i}^{O}}{K}\right] \\
+g_{B}\left[ q_{B}^{O}H^{I}\left( 1-s\right) \frac{n\left(
1-\sum_{i}g_{i}q_{i}^{O}\right) }{K}\right] +g_{A}\left[ \left(
1-q_{A}^{O}\right) H^{O}\left( 1-s\right) \frac{n\sum_{i}g_{i}q_{i}^{O}}{K}%
\right] \\
=\left[ \left( g_{H}q_{H}^{O}+g_{B}q_{B}^{O}\right) H^{I}\left( 1-s\right)
\left( 1-\sum_{i}g_{i}q_{i}^{O}\right) +\left( g_{H}\left(
1-q_{H}^{O}\right) +g_{A}\left( 1-q_{A}^{O}\right) \right) H^{O}\left(
1-s\right) \sum_{i}g_{i}q_{i}^{O}\right] \frac{n}{K}.
\end{gather*}

\section*{Appendix 2 Derivation of selection dynamics}

Selection is described by equation

\begin{equation}
\dot{g}_{s}=g_{s}\left[ \left( \bar{\Phi}_{s}(q_{s}^{O})-\bar{\Phi}%
(g,q)\right) -\left( \bar{D}_{s}(g,q,n)-\bar{D}(g,q,n)\right) -\left( \bar{%
\Psi}_{s}(q_{s})-\bar{\Psi}(g,q)\right) \right] ,
\end{equation}%
\bigskip

where for strategy $s$ average fertility is equal to $\bar{\Phi}%
_{s}(q_{s}^{O})=q_{s}^{O}\Phi $ and background mortality is%
\begin{equation*}
\bar{\Psi}_{s}(q_{s})=q_{s}^{O}\Psi _{O}+\left( 1-q_{s}^{O}\right) \Psi _{I}.
\end{equation*}%
Then%
\begin{equation*}
\bar{\Phi}(g,q)=\sum_{i}g_{i}\bar{\Phi}_{i}(q_{i})=\Phi
\sum_{i}g_{i}q_{i}^{O},
\end{equation*}%
\begin{equation*}
\bar{\Psi}(g,q)=\sum_{i}g_{i}\bar{\Psi}_{i}(q_{i})=\Psi
_{O}\sum_{i}g_{i}q_{i}^{O}+\Psi _{I}\left( 1-\sum_{i}g_{i}q_{i}^{O}\right) .
\end{equation*}%
Then we can derive the bracketed terms: 
\begin{gather*}
\left( \bar{\Phi}_{s}(q_{s})-\bar{\Phi}(g,q)\right) -\left( \bar{\Psi}%
_{s}(q_{s})-\bar{\Psi}(g,q)\right) =\left[ q_{s}^{O}\Phi -\Phi
\sum_{i}g_{i}q_{i}^{O}\right] \\
-\left[ q_{s}^{O}\Psi _{O}+\left( 1-q_{s}^{O}\right) \Psi _{I}-\Psi
_{O}\sum_{i}g_{i}q_{i}^{O}-\Psi _{I}\left( 1-\sum_{i}g_{i}q_{i}^{O}\right) %
\right] \\
=\left( q_{s}^{O}-\sum_{i}g_{i}q_{s}^{O}\right) \left[ \Phi -\Psi _{O}+\Psi
_{I}\right] .
\end{gather*}%
Therefore, selection equation can be presented as%
\begin{equation}
\dot{g}_{s}=g_{s}\left[ \left[ \Phi -\Psi _{O}+\Psi _{I}\right] \left(
q_{s}^{O}-\sum_{i}gq_{i}^{O}\right) -\left( \bar{D}_{s}(g,q,n)-\bar{D}%
(g,q,n)\right) \right] .
\end{equation}

\end{document}